\documentclass{aa}

\usepackage[desactivate]{linenoaa}

\usepackage{graphicx}
\usepackage{txfonts}
\usepackage[colorlinks=true, linkcolor=blue, citecolor=blue, filecolor=blue, urlcolor=blue]{hyperref}
\usepackage{bm}

\usepackage{xcolor}
\usepackage{lscape}
\usepackage{multirow}

\usepackage{xspace}

\newcommand{\project}[1]{\textsl{#1}}

\newcommand{\Spitzer}{\project{Spitzer}}

\begin{document}

\title{Extended source fringe flats for the JWST MIRI Medium Resolution Spectrometer}

\titlerunning{Extended source fringe flats for the JWST MIRI MRS}

\authorrunning{N. Crouzet et al.}

\author{
N. Crouzet\inst{1}
\and
M. Mueller\inst{2}
\and
B. Sargent\inst{3}
\and
F. Lahuis\inst{4}
\and
D. Kester\inst{4}
\and
G. Yang\inst{5}
\and
I. Argyriou\inst{6}
\and
D. Gasman\inst{6}
\and
P. J. Kavanagh\inst{7}
\and
A. Labiano\inst{8,9}
\and
K. Larson\inst{10}
\and
D. R. Law\inst{3}
\and
J. Álvarez-Márquez\inst{9}
\and
B. R. Brandl\inst{1,11}
\and
A. Glasse\inst{12,13,14}
\and
P. Patapis\inst{15}
\and
P. R. Roelfsema\inst{4}
\and
Ł. Tychoniec\inst{1}
\and
E. F. van Dishoeck\inst{1}
\and
G. S. Wright\inst{14}
}

\institute{
Leiden Observatory, Leiden University, P.O. Box 9513, 2300 RA Leiden, The Netherlands \\
\email{crouzet@strw.leidenuniv.nl}
\and
Kapteyn Astronomical Institute, Rijksuniversiteit Groningen, Postbus 800, 9700 AV Groningen, The Netherlands
\and
Space Telescope Science Institute, 3700 San Martin Drive, Baltimore, MD 21218, USA
\and
SRON Netherlands Institute for Space Research, Postbus 800, 9700 AV Groningen, The Netherlands
\and
Nanjing Institute of Astronomical Optics \& Technology, Chinese Academy of Sciences, Nanjing 210042, China
\and
Institute of Astronomy, KU Leuven, Celestijnenlaan 200D, 3001 Leuven, Belgium
\and
Department of Physics, Maynooth University, Maynooth, Co. Kildare, Ireland
\and
Telespazio UK for the European Space Agency, ESAC, Camino Bajo del Castillo s/n, 28692 Villanueva de la Ca\~nada, Spain
\and
Centro de Astrobiología (CAB), CSIC-INTA, Ctra. de Ajalvir km 4, Torrejón de Ardoz, E-28850, Madrid, Spain
\and
AURA for the European Space Agency (ESA), Space Telescope Science Institute, 3700 San Martin Drive, Baltimore, MD 21218, USA
\and
Faculty of Aerospace Engineering, Delft University of Technology, Kluyverweg 1, 2629 HS Delft, The Netherlands
\and
Institute for Astronomy, University of Edinburgh, Royal Observatory, Blackford Hill, Edinburgh, EH9 3HJ, UK
\and
Centre for Exoplanet Science, University of Edinburgh, Edinburgh, EH9 3HJ, UK
\and
UK Astronomy Technology Centre, Royal Observatory Edinburgh, Blackford Hill, Edinburgh EH9 3HJ, UK
\and
Institute of Particle Physics and Astrophysics, ETH Zurich, Wolfgang-Pauli-Strasse 27, 8093, Zurich, Switzerland
}

\date{Received ...; accepted ...}

\abstract
{The detectors of the JWST Mid-Infrared Instrument (MIRI) Medium Resolution Spectrometer (MRS) form low-finesse resonating cavities that cause periodic count rate modulations (fringes) with peak amplitudes of up to 15\% for sources external to MIRI.  To detect weak features on a strong continuum and reliably measure line fluxes and line-flux ratios, fringe correction is crucial.}
{This paper describes the first of two steps implemented in the JWST Science Calibration Pipeline, which is the division by a static fringe flat that removes the bulk of the fringes for extended sources.}
{Fringe flats were derived by fitting a numerical model to observations of spatially extended sources.  The model includes fringes that originate from two resonating cavities in the detector substrate (a third fringe component that originates from the dichroic filters is not included). The model, numerical implementation, and resulting fringe flats are described, and the efficiency of the calibration was evaluated for sources of various spatial extents on the detector.}
{Flight fringe flats are obtained from observations of the planetary nebula NGC 7027. The two fringe components are well recovered and fitted by the model. The derived parameters are used to build a fringe flat for each MRS spectral band, except for 1A and 1B due to the low signal-to-noise ratio of NGC 7027 in these bands. When applied to extended sources, fringe amplitudes are reduced to the sub-percent level on individual spaxels. For point sources, they are reduced to amplitudes between 1 and 5\% considering individual spaxels and a single dither position, and decrease to the 1 to 2\% level after two-dimensional residual fringe correction.}
{The fringe flats derived from this work are the reference files currently in use by the JWST Science Calibration Pipeline. They provide an efficient calibration for extended sources, and are less efficient for point sources. Future improvements of these fringe flats are possible. The fringe modelling method could also be tested on individual semi-extended or point sources.}


\maketitle

\section{Introduction} \label{sec:intro}

\begin{figure*}[ht!]
    \includegraphics[width=\columnwidth]{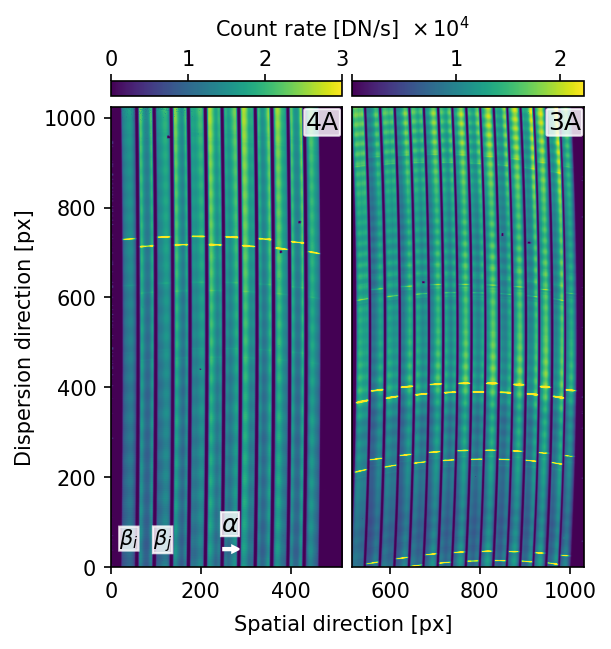}
    \includegraphics[width=\columnwidth]{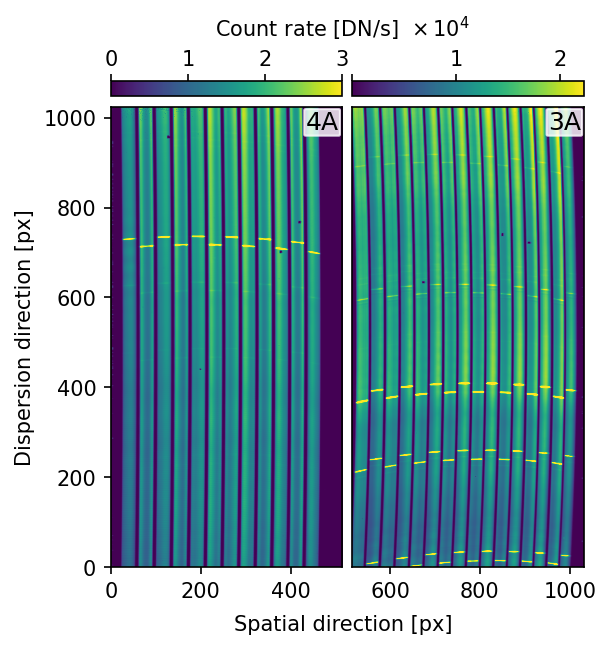}
    \caption{Left: MIRI MRS detector image of the spatially extended source NGC 7027 for the long wavelength detector MIRIFULONG, sub-band A, after slope fitting (level 1 `\texttt{rate}' output of the JWST calibration pipeline). Spectral bands 4A and 3A are on the left and right hand sides, respectively, each with its own colour scale indicated at the top. The two half-detectors are split for clarity. Fringes appear along the dispersion (vertical) direction. The two spatial coordinates are illustrated: $\alpha$ varies within each slice perpendicularly to the slice as indicated by the arrow, $\beta$ is constant for each slice with $i$ and $j$ denoting the slice number. The conspicuous bright features at specific wavelengths (that appear along the horizontal, spatial direction) are bright emission lines from the source. Right: Same image after division by the fringe flat derived from that source, with the same colour scales as in the left panel. Fringes are largely suppressed.
    }
    \label{fig:rawFringes}
\end{figure*}

\begin{figure}[h!]
    \includegraphics[width=0.95\columnwidth]{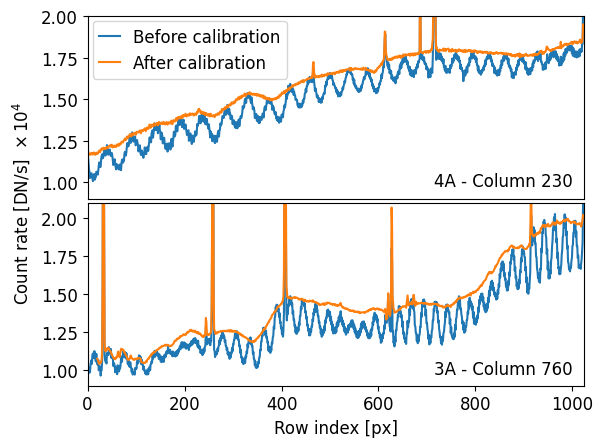}
    \caption{Vertical cuts along columns 230 (band 4A, top) and 760 (band 3A, bottom) of NGC~7027 images for the MIRIFULONG detector, sub-band A, before (blue) and after (orange) calibration by the flatfield and by the fringe flat derived from that source, as done in the JWST calibration pipeline. Each column is a spectrum in detector space, neglecting slice curvature and distortion (the spectrum of a fixed point in spatial coordinates would follow an iso-$\alpha$ line). The fringes are the strong periodic ripples in the spectral baseline. The peaks are emission lines from the source. Each fringe flat is normalised to have a maximum equal to unity, so the fluxes after fringe flat calibration follow roughly the fringe maxima (photometric calibration is performed in a later step in the JWST calibration pipeline).
    }
    \label{fig:columnCut-fringecor}
\end{figure}

The JWST, launched on 25 December 2021, provides unprecedented sensitivity for astronomical observations at near- and mid-infrared wavelengths \citep{Gardner2006, Gardner2023}.
The Mid-Infrared Instrument \citep[MIRI;][]{rieke2015, Wright2023} covers the 5 to 28~$\mu$m range and includes the Medium Resolution Spectrometer \citep[MRS;][]{Wells2015}.
This paper describes the \texttt{fringe} calibration step for MIRI MRS that is included in the JWST Science Calibration Pipeline \citep{Bushouse2019, Bushouse2020, Bushouse2025}. This calibration was derived from sources external to MIRI.
As is common for infrared spectrometers, the MIRI MRS suffers from fringing, periodic count rate modulations due to standing waves in the optical path. In the case of the MRS, the main resonating cavities are in the detector itself. Depending on wavelength, MRS fringing varies in peak amplitude between 2\% and 15\% of the spectral baseline for sources external to MIRI (Figs. \ref{fig:rawFringes} and \ref{fig:columnCut-fringecor}) and up to 20\% for the MRS internal calibration source \citep{20ArRiRe}. 
This severely limits the detection of weak diagnostic features both in absorption and emission superposed on strong continuum emission. Examples include molecular lines in protostellar sources
\citep[e.g.][]{Lahuis2000,vanGelder2024}, protoplanetary disks \citep[e.g.][]{Carr2008,Salyk2011,Grant2023,Banzatti2023,Gasman2023b,Pontoppidan2024,Temmink2024}, and active galactic nuclei \citep[e.g.][]{Lahuis2007,Buiten2025}.
The determination of molecular abundances, the inferred carbon to oxygen ratio, and isotope ratios in the atmospheres of exoplanets and brown dwarfs can be affected \citep[e.g.][]{Barrado2023, Kuhnle2024, Deming2024, ZhangY2021, ZhangY2024, GonzalezPicos2024}.
Some of these molecules have regular patterns that could be misinterpreted as fringing in automatic procedures, thus requiring particular caution
\citep{Gasman2023a}. Also, ice bands due to minor species that often carry the most interesting chemical information typically absorb only a few percent of the continuum radiation \citep[e.g.][]{Rocha2024,Brunken2024,Slavicinska2025}. Observations therefore need to reach a signal-to-noise ratio ($S/N$) of a few hundred on the continuum to fully retrieve the physical and chemical information from the data. 

In many astronomical sources, from nearby photodissociation regions to distant galaxies, the ratios of ionic (e.g. [Ne II], [Ne III]) or molecular (e.g. H$_2$) emission lines are often used to probe the physical conditions of the gas, such as density, temperature, or excitation mechanism. The effect of uncorrected fringes, which can boost or reduce the measured line flux due to constructive or destructive interference, can make this approach extremely unreliable.
For example, let us assume two intrinsically equally strong emission lines [Ne II] and [Ne III], and a realistic, uncorrected fringe amplitude of 20\%. In the case that [Ne II] happens to be detected at a fringe maximum and [Ne III] at a fringe minimum, the ratio [Ne III]/[Ne II] is 0.67. In the opposite case, the ratio [Ne III]/[Ne II] is 1.5, in other words, more than two times larger. This is the worst-case scenario, and in most cases the uncertainty is less, but still too large for any reliable line modelling.

Different spectrometers require different fringe-removal techniques.  Spectra from the \Spitzer's Infrared Spectrograph (IRS), for example, were de-fringed by fitting sinusoids to extracted spectra \citep{Lahuis2003}, following similar work for the  Infrared Space Observatory \citep[ISO,][]{Kester2003}.  For an imaging spectrometer such as MIRI MRS, fringes should be corrected across the field of view.
The JWST Science Calibration Pipeline\footnote{\url{https://jwst-pipeline.readthedocs.io/en/stable/}} contains two steps for the correction of fringes for MIRI MRS, which are implemented in the \verb|Spec2Pipeline|\footnote{\url{https://jwst-docs.stsci.edu/jwst-science-calibration-pipeline/stages-of-jwst-data-processing/calwebb_spec2}}\textsuperscript{,}\footnote{\url{https://jwst-pipeline.readthedocs.io/en/stable/jwst/pipeline/calwebb_spec2.html}}. The first is the division by a static fringe flat and is performed by the \verb|fringe| step\footnote{\url{https://jwst-pipeline.readthedocs.io/en/stable/jwst/fringe/index.html}}. Fringe flats are derived from high $S/N$ observations of spatially extended sources (with spatially uniform illumination across the field of view). The derivation of in-flight fringe flats is described in this paper. Then, residual fringes are corrected by the \verb|residual_fringe| step, which attempts to remove remaining fringes at the two-dimensional (2D) detector level through an empirical sinusoid fitting method. A one-dimensional (1D) residual fringe correction is also available as an option during 1D spectral extraction from the data cubes. The residual fringe correction procedure will be described in a forthcoming paper (Kavanagh et al. in prep.).
The nature of the fringes and mitigation methods for point sources are described in \citet{Argyriou2020}, \citet{Gasman2023a}, \citet{Pontoppidan2024}, and \citet{Gasman2025}, they have been proven to work well and yield residuals at the sub-percent level \citep{Gasman2023a, Gasman2023b, Grant2023}. However, they are not well suited to extended sources (and vice versa) because their illumination pattern differs significantly. \citet{Argyriou2020} showed that the fringe pattern for point sources is highly dependent on the incidence angle of the source and the sampling of the incoming wavefront, that is the point spread function (PSF). In contrast, the fringe pattern for extended sources is the result of optical interference from a continuum of incidence angles and source positions.

To derive extended source fringe flats, a numerical model was employed to fit the fringes to the continuum portion of the data as a function of wavelength.  As discussed below, the model contained two components with slightly different periods that correspond to two resonating cavities within the MIRI detectors \citep{20ArRiRe}.
The model employed a Bayesian algorithm to fit model parameters and allowed smooth changes of, for example, detector thickness and material optical properties as a function of projected wavelength across the detector (consistent with ground knowledge and testing). This model is presented in Sect.\ \ref{sec:model}.
Extended source fringe flats were successively derived from observations obtained during the MIRI Flight Model (FM) ground test campaign, from in-flight observations of the Cat's Eye Nebula (NGC~6543) obtained during commissioning, and from in-flight observations of the Jewel Bug Nebula (NGC~7027) obtained during Cycle~1. These observations are described in Sect. \ref{sec:observations}.
The fringe flats obtained from applying the numerical model to these observations are described in Sect. \ref{sec:results}. The efficiency of the fringe correction is evaluated on various sources in Sect.\ \ref{sec:efficiency}. A discussion and possible improvements are presented in Sect. \ref{sec:discussion}, and conclusions are given in Sect. \ref{sec:conclusion}.

\section{Fringe modelling}
\label{sec:model}

\subsection{Overview of the MRS}
\label{sec:MRS}

Light entering the MRS is split into four separate spectral ``channels'' (channels 1 to 4 from shortest to longest wavelengths) using dichroic filters.  For each channel, the MRS contains an integral field unit (IFU). Photons that enter an IFU first encounter the image slicer that separates the spatial field of view of each channel into a number of thin slices. These are then spectrally dispersed using a diffraction grating and the wavelengths and spatial positions are projected onto the detector. The MRS has two 1k$\times$1k pixel Si:As impurity band conduction detector arrays \citep{rieke2015_detectors} that can operate simultaneously: MIRIFUSHORT employed by channels 1 and 2 ($4.83-11.87\,\mu$m) and MIRIFULONG employed by channels 3 and 4 \citep[$11.47-28.82\,\mu$m;][]{Wells2015}. 
Each channel is spectrally divided into three sub-bands A, B, and C (or SHORT, MEDIUM, LONG); obtaining a full spectrum therefore requires three exposures.  When changing sub-bands, different dichroics and diffraction gratings are rotated into the optical path; image slicers and detectors stay the same. 
On the detector, slices are spectrally dispersed in the vertical direction and are roughly parallel to one another. Fig.\ \ref{fig:rawFringes} shows detector-level MRS data of NGC 7027 where fringes can be visualized and Fig.\ \ref{fig:columnCut-fringecor} shows a cut of individual columns (observations and data processing are detailed in Sect. \ref{sec:observations}). Each channel occupies one half of a detector; note that neighbouring slices on the detector are not neighbours on sky.  The wavelength direction, $\lambda$, follows the detector $y$ axis, with some amount of distortion and curvature that varies between slices and channels. Within a slice, the detector $x$ axis is roughly aligned with the spatial dimension along the image slicer $\alpha$ (the iso-$\alpha$ lines follow the slice curvature).  The spatial direction perpendicular to the slices on the IFU image slicer, $\beta$, is constant within a slice but differs from slice to slice \citep[see Fig. 3 of][]{Argyriou2020}.
Fringes manifest themselves as a quasi-oscillation on the vertical axis that is roughly aligned with wavelength.

To convert raw data to calibrated spectra, three pipeline levels are run \citep[see][for a description of the MRS pipeline]{Labiano2016}.  In a first level, for each detector pixel the average rate of charge build-up is determined (slope fitting).  Level 2 uses these derived slope images as input, assigns to each pixel its position on the sky and wavelength, and applies the spectrophotometric correction factors to extract the astrophysical flux values.  In level 3, calibrated detector images from different dither positions and associated background observations are combined onto a regular three-dimensional (3D) grid.  The calibrated target spectrum is extracted at the end of level 3 from the final regularised data cubes. 
Standard-pipeline fringe correction occurs at level 2: slope images are divided by a fringe flat and the residual fringe correction operates on these corrected detector images. 
An alternative (or supplementary) 1D residual fringe correction can be performed on the 1D spectrum extracted from the level 3 spectral cubes.

For a more detailed, general description of the MRS, the reader is referred to \citet{Wells2015}; fringes and how they are related to the MRS detectors are described in \citet{20ArRiRe} and \citet{Argyriou2020}. An overview of the science performance of JWST and its science instruments during commissioning is given in \citet{Rigby2023}, an overview of early in flight performance of MIRI is presented in \citet{Wright2023}, and in-flight performance of MIRI MRS specifically can be found in \citet{Argyriou2023}.

\begin{figure}
    \centering
    \includegraphics[width=0.9\columnwidth]{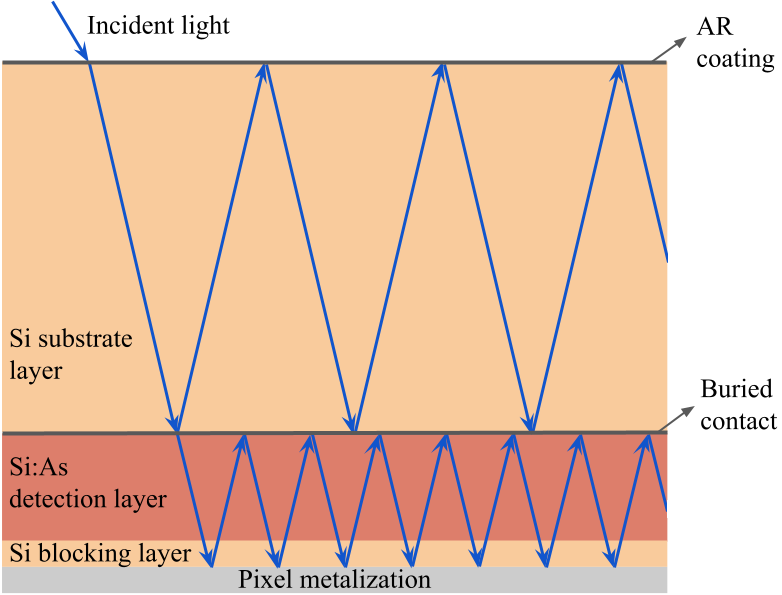}
    \caption{Physical origin of spectral fringes. Photons reflected within the resonating cavities of the MRS short and long wavelength detectors undergo constructive or destructive interference depending on wavelength. The two cavities consist of the AR coating to the buried contact and the buried contact to the pixel metalization.}
    \label{fig:physics}
\end{figure}

\subsection{Physical origin of fringes}
\label{sec:physics}

The origin of spectral fringes is coherent reflection within the optical path, where cavities act as Fabry-Pérot etalons.  Interference between wavelengths that were reflected $n$ times creates the typical sensitivity modulations (maxima for constructive interference, minima for destructive interference) that are periodic with wavenumber. Fig.\ \ref{fig:physics} illustrates the detector geometry giving rise to interference.
In the case of MIRI MRS, the most relevant resonating cavities are within the detectors themselves and the dichroic filters.  As demonstrated in \citet{20ArRiRe}, there are two resonating cavities in the detectors: one is defined by the anti-reflection (AR) coating and the buried contact, the other by the buried contact and the pixel metalization.
At short wavelengths ($\lambda\,<\,10\,\mu m$), the buried contact is almost transparent and the infrared-active detection layer absorbs the smallest fraction of incoming wavelengths \citep{gaspar2021}. This results in one dominant fringe pattern with a frequency proportional to the optical thickness bound by the AR coating and the pixel metalization. At intermediate wavelengths ($10\,\mu m\,<\lambda\,<\,18\,\mu m$), the buried contact becomes gradually more reflective while the infrared-active detection layer still does not absorb all incoming photons. In this regime, a strong beating between the fringes produced by the AR coating to buried contact and buried contact to pixel metalization is measured. This beating makes fringe modelling significantly more challenging. Finally, at long wavelengths ($\lambda\,>\,18\,\mu m$), the buried contact is at its most reflective, and the infrared-active detection layer absorbs the largest fraction of incoming photons. The dominant fringe is then defined by the cavity between the AR coating and the buried contact. It can be noted that the AR coating thickness is optimised for 6 and 16 $\mu$m for the short and long wavelength detector respectively \citep{rieke2015_detectors}, which also has an impact on the amplitude of the fringes.

A third fringe component of lower amplitude (up to $\sim 5\%$ peak-to-peak) and higher frequency appears in long wavelength detector data ($11.47-28.82\;\mu$m). It is attributed to a resonating cavity in the MRS dichroics (the observed fringe period is consistent with the known optical thickness of the dichroics), upstream of the detector, echelle grating, IFU optics, and anamorphic pre-optics \citep{Wells2015}. This fringe component is not modelled in the fringe flats since it is a complex fringe product of more than one dichroic in the optical path.

\subsection{Fringe modelling}
\label{sect:math}

For monochromatic light that impinges on the detector at a well-defined incidence angle, the fringe pattern is described by the well-known Airy formula describing general Fabry-Pérot etalons. 
The transmittance $T(w)$ of each MIRI detector cavity is proportional to:
\begin{equation}
\label{eq:Airy}
T(w) \propto
\frac{1}{1+F\sin^2\left(\pi w / P + \phi \right)}
\end{equation}
with coefficient of finesse $F$ (named finesse hereafter), period $P$ (in wavenumber space), wavenumber $w$, and an empirical phase shift $\phi$. ``Transmittance'' is a general term for Fabry-Pérot etalons, although here the light is absorbed (converted to photo-electrons) by the infrared-active layer of the detector.
The finesse controls the fringe amplitude and is controlled by the reflectivity of the resonating cavity; it is fairly low for the MRS detectors as compared to custom-made high-finesse etalons \citep[such as the one used for MRS wavelength calibration on the ground,][]{Labiano2021} but varies noticeably with wavelength, even across a single MRS slice. 
For low finesse, the Airy formula is well approximated by a sinusoidal function, the leading term in the Taylor expansion; it is under this assumption, of small fringe amplitudes, that the residual fringe correction employs sinusoids.  With increasing finesse, the fringe pattern becomes noticeably non-sinusoidal.  

The fringe frequency $f$ (that is $1/P$)  is determined by the optical thickness of the resonating cavity: 
\begin{equation}
\label{eq:period}
    f = 2 n l \cos\theta    
\end{equation}
with geometric thickness $l$, index of refraction $n$, and refraction angle $\theta$ (the angle the light travels through the etalon, which is related to the incidence angle by the Snell-Descartes relation).  The fitted fringe frequency is very well reproduced across different fits to MRS data (but is specific to each channel and band, since the index of refraction is wavelength dependent). 
The fringe period and phase should, mathematically speaking, be deterministic given the detector material, geometry, and a fixed incidence angle of the incoming photons.  In practice, however, variations in material refractive coefficients with wavelength, variations in geometric thickness of the order of micrometers across the detector chip surface, and variations in incidence angle for different illumination patterns make the effort of modelling these parameters extremely degenerate and complex. For these reasons, the period (or frequency) and the phase must be fitted to the data.  The underlying physical reason is that the cavity is in high fringe order: the argument to the sine square in the Airy formula in Eq.~(\ref{eq:Airy}) is much larger than its $\pi$ periodicity, such that even subtle changes in effective wavelength or incidence angle cause significant shifts in fringe phase.

\subsection{Point sources versus extended sources}
\label{sec:Point sources versus extended sources}

\begin{figure}
    \centering
    \includegraphics[width=0.9\columnwidth]{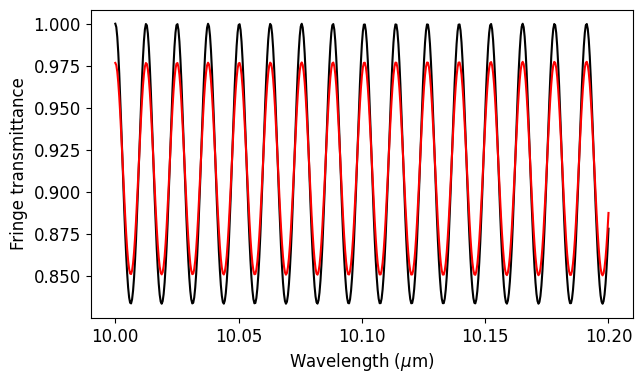}
    \caption{Comparison of two synthetic fringe patterns: black for a perfectly collimated source, where the fringe pattern follows the Airy formula in Eq.~(\ref{eq:Airy}), and red for a source with a small range in effective period ($P/\Delta P  = 2000$) that mimics an extended source. The resulting reduction in effective fringe amplitude is readily apparent.}
    \label{fig:washingOut}
\end{figure}

The mathematical description based on Eqs.~\ref{eq:Airy} and \ref{eq:period} is valid for perfectly collimated or monochromatic light.  Photons that hit the MRS detectors are neither: for each detector pixel, there is a distribution in wavelength and incidence angle, albeit small but significant.  The resulting fringe pattern is thus a convolution over a range of fringe periods. 
Fringe flats used in the \verb|Spec2Pipeline| for MRS are based on model fits to observations of spatially extended sources. Extended source observations result in a wider set of incidence angles on the detector pixels, so the fringe pattern is averaged over a larger range of periods and the fringe amplitude is smaller. Point source observations result in a narrower set of incidence angles on the detector pixels, so the fringe pattern is averaged over a smaller range of periods and the fringe amplitude is larger. An additional consideration is that for point sources the mean incidence angle and the range of sampled angles can change across the PSF, which results in a dependence of the fringe amplitude and phase on the part of the PSF that is sampled \citep{Argyriou2020}. Fig.~\ref{fig:washingOut} illustrates the partial cancelling of the fringe pattern for a larger set of incidence angles, hence periods, which results in a reduced effective fringe amplitude. The main consequence is that a fringe flat derived from an extended source observation and detector illumination cannot lead to a perfect fringe correction for point sources.

The residual fringe contrast in final extracted spectra of point sources after applying fringe flat and residual fringe corrections was found to be at the percent level or better, sufficient for many science cases \citep{Argyriou2023}.
Observers requiring higher fidelity fringe removal need a different fringe correction method.  One possible approach is to use a feature-poor point source as empirical fringe calibrator. Given the sensitive dependence of the fringe pattern on position offsets, target acquisition would need to be activated for both observations and the same dither pattern would need to be used to ensure that the source is located at the same position on the detector as the calibrator. Commissioning results (Sect. \ref{sec:observations}) and further studies by \citet{Gasman2023a} and \citet{Pontoppidan2024} indicate that point source data can indeed be corrected for fringes at high fidelity using an empirical point source fringe calibrator such as a hot (O- or A-type) star or asteroid. As a limitation, if the point source is not located at the same position as the calibrator, another fringe correction method is required.

\subsection{Numerical fringe model}
\label{sec:numerical model}

\begin{figure}
    \centering
    \includegraphics[width=0.9\columnwidth]{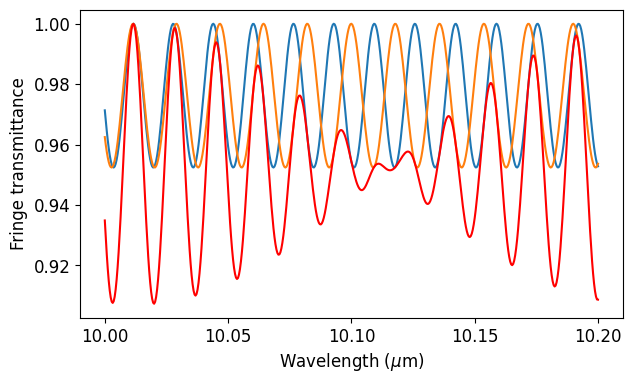}
    \caption{Synthetic fringe pattern for a perfectly collimated source that enters two resonating cavities, which mimics the two fringe components (blue and orange). Each of them follows the Airy formula in Eq.~(\ref{eq:Airy}) with its own period, which are close to each other, and both have the same finesse. Their combination produces beating (red).}
    \label{fig:beating}
\end{figure}

\begin{figure}
    \centering
    \includegraphics[width=\columnwidth]{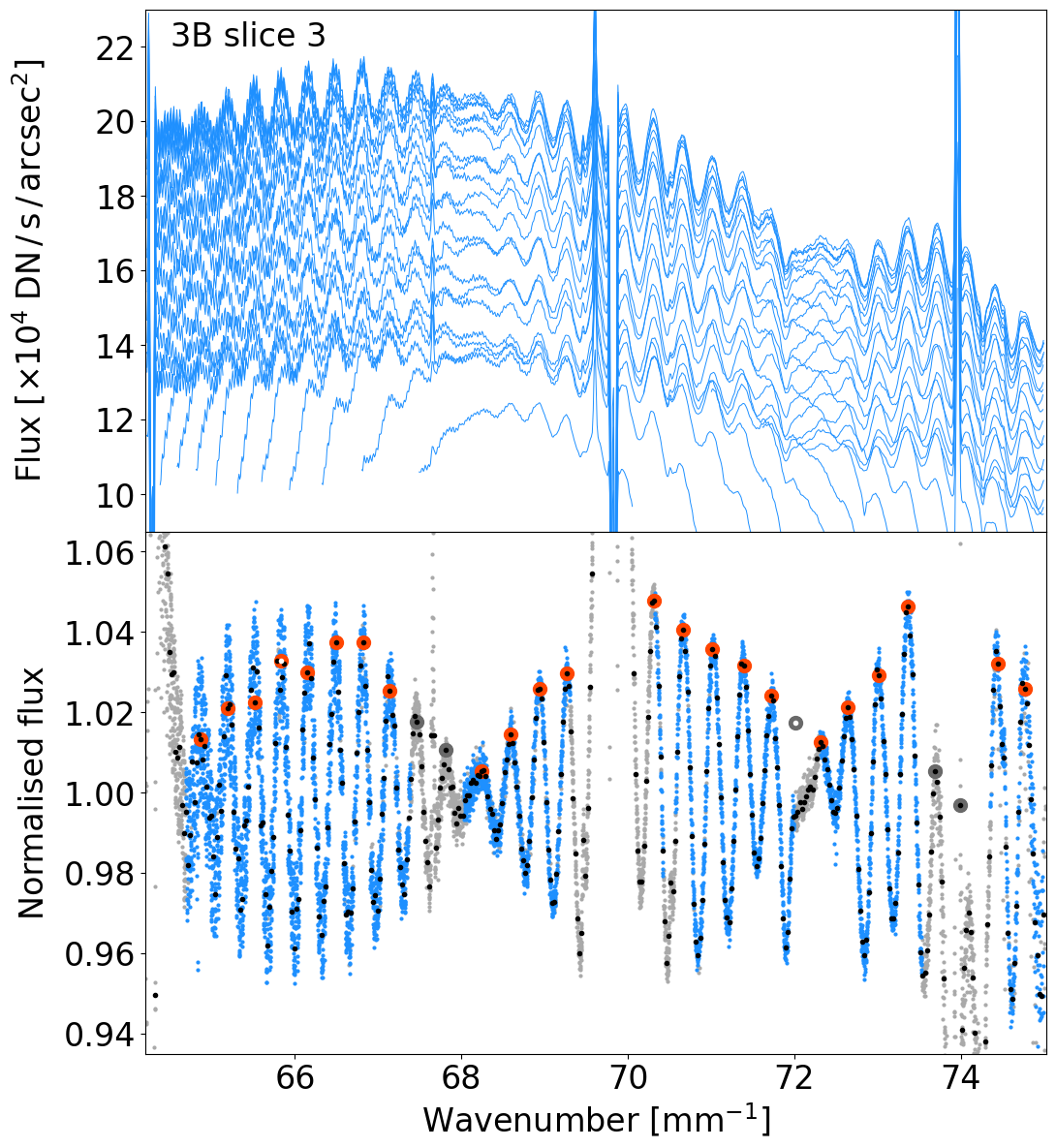}
    \caption{Flux normalisation and search for local maxima for band 3B, slice 3 for the Cat's Eye Nebula data. Top: Flux for all columns of the slice measured from the `\texttt{rate}' files. Spikes seen on all columns are spectral lines from the source. Bottom: Normalised flux after removing the continuum for each column (blue points), masked regions (grey points), normalised flux after binning (black points), and local maxima (red circles) including those that were initially missed (white points) or were discarded (grey circles).}
    \label{fig:flux-wtops-3B-slice3}
\end{figure}

The observed data were fitted using a numerical model composed of a smooth continuum and two fringe components of slightly different period and amplitude. 
The double periodicity is clearly seen in the data as a beating pattern, especially around the central wavelengths of the MRS range (see Fig.~\ref{fig:beating} for simulated beating and Fig.~\ref{fig:flux-wtops-3B-slice3} for observed beating in MRS data).
Fringe models were implemented using the \verb|EtalonModel| class in the \verb|BayesicFitting| package\footnote{\url{https://bayesicfitting.nl/}} \citep{Kester2021}, which is included in the JWST science calibration pipeline. Model parameters for a single fringe component were the finesse, the frequency, and the phase. 
Each slice of each detector and band was fitted separately.  Model parameters that described the continuum, the fringe frequencies, and the finesse of one fringe component were allowed to vary smoothly with wavenumber but were assumed to be constant with $\alpha$ (the finesse of the second fringe component was assumed to be constant with wavenumber).  Each fringe model was given a phase parameter that was also fitted to the data.
The same numerical model was used pre-launch to fit the fringe pattern observed in ground-test observations.

\begin{figure}
    \centering
    \includegraphics[width=\columnwidth]{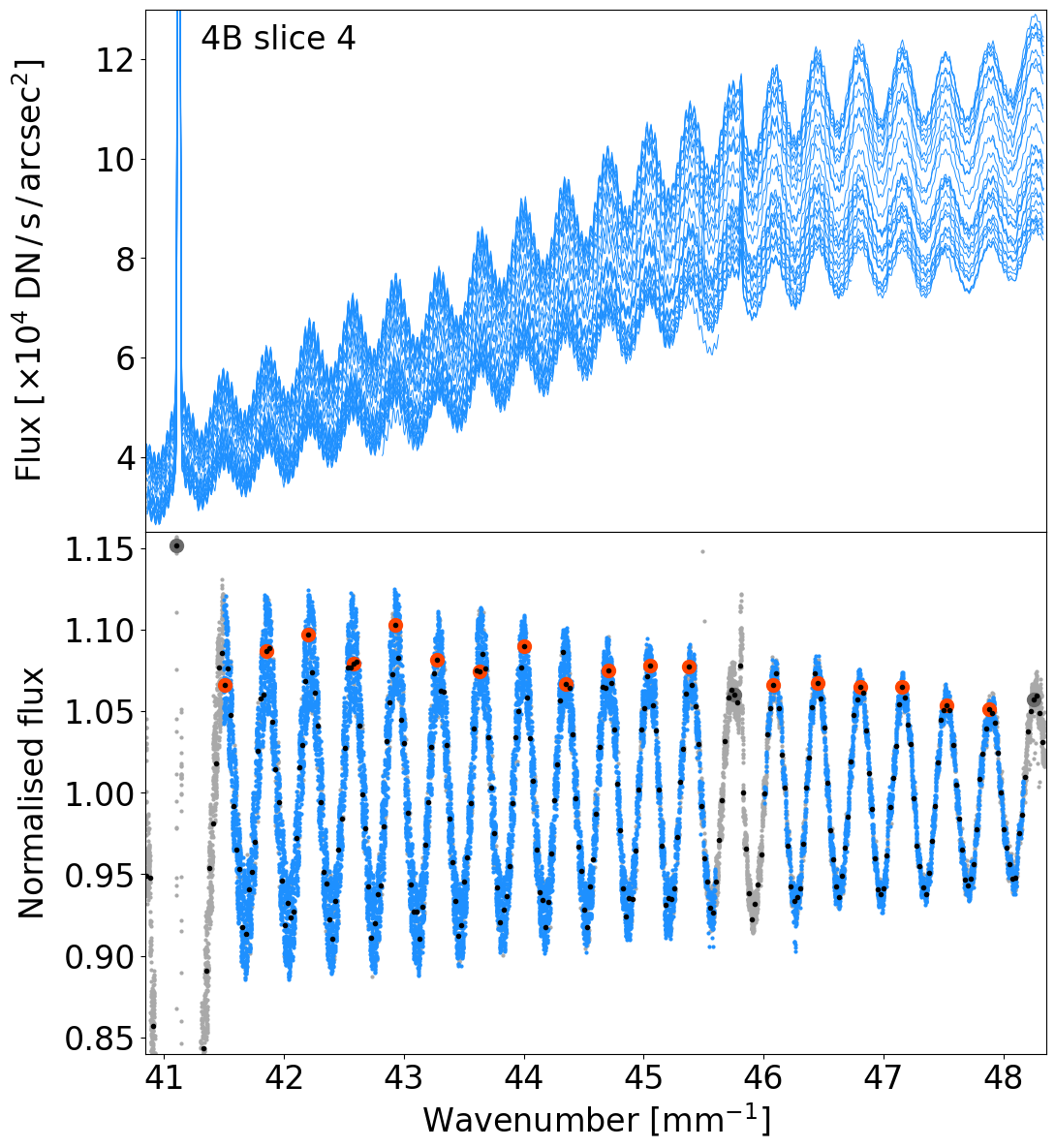}
    \caption{Same as Fig.~\ref{fig:flux-wtops-3B-slice3} for band 4B, slice 4. The continuum in the top panel has less curvature than in band 3B (Fig.~\ref{fig:flux-wtops-3B-slice3}) because the slices are straighter on the detector (less distortion) in channel 4 compared to channel 3.}
    \label{fig:flux-wtops-4B-slice10}
\end{figure}

For more robust numerical modelling, data were normalised prior to fitting.  A first-pass continuum was computed for each column of each slice and divided out, which led to flux values centred on unity (Figs.~\ref{fig:flux-wtops-3B-slice3} and \ref{fig:flux-wtops-4B-slice10}).  Wavelengths were converted to wavenumbers, then shifted to a minimum value of zero for simplicity (without loss of generality, albeit unphysical).
Pixel flux values suffered from a low-amplitude (but clearly discernible) apparent offset with a periodicity of two rows (MIRIFUSHORT) or four rows (MIRIFULONG).  The former was an even/odd row effect \citep{Morrison2023}, the latter was significant on MIRIFULONG and is thought to be due to fringes in the dichroic filters.
The current fringe flats were not designed to correct for such high-frequency effects. To mitigate this issue, data were smoothed via convolutions along the dispersion axis to remove these variations before fringe fitting. This was found to be necessary to obtain a useful fit.

\begin{figure}
    \centering
    \includegraphics[width=\columnwidth]{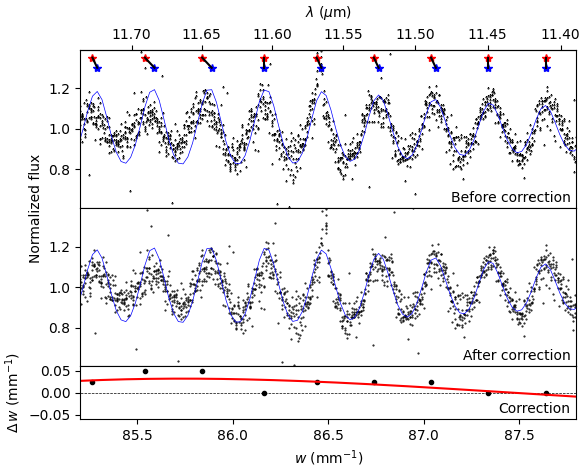}
    \caption{Wavelength recalibration for band 2C, slice 8, zoomed in a small wavelength region. $\lambda$ is the wavelength, $w$ is the wavenumber, and $\Delta w$ is the wavenumber difference. The top panel shows the normalised flux before wavelength recalibration (black dots) and the numerical fringe model (blue line). The red and blue stars indicate the local maxima of the data and fringe model, respectively, and the black lines illustrate their shifts in wavelength. The middle panel shows the normalised flux after wavelength recalibration (black dots) and the same fringe model as in the top panel (blue line). The bottom panel shows the wavenumber differences between the  local maxima of the data and of the fringe model (black points) and the spline function to model and correct for these (red line).}
    \label{fig:wavelength recalibration}
\end{figure}

MRS data were found not to follow a perfectly periodic fringe pattern as described above.  To fit the data, the argument to the sine in the Airy formula must be allowed to vary slightly with wavelength. This proved necessary for the fit to converge on a stable fringe solution.
Physically, this can be thought of as either a slight recalibration of the wavelength (or wavenumber) or as a dependence of optical thickness on detector position.  Either could be true, as could be a combination of both (the two are degenerate). This recalibration was purely internal to the fringe fitting procedure. Thus we did not use any ``wavelength recalibration'' obtained during fringe fitting to glean information on actual, physical wavelength calibration. The resulting fringe flats are inherently in the detector-pixel space, independently of any wavelength calibration. Thus it is not expected that any future updates to the wavelength calibration necessitate an updated fringe flat.
Algorithmically, the wavelength recalibration consisted of numerically identifying local flux maxima, which can be complicated in the presence of noise and spectral features, especially in areas of low fringe amplitude due to beating (Fig.~\ref{fig:flux-wtops-3B-slice3} and \ref{fig:flux-wtops-4B-slice10}). This was done by binning the normalised flux using 40 bins per wavenumber, fitting a second order polynomial to each group of five consecutive points, and testing if the polynomial fit value of the middle point was larger than those of its four neighbours. Then, missing maxima were searched for and inserted.
Wavelength regions affected by spectral features were identified manually and masked out. Then, a smooth, low-order spline wavenumber correction was applied to the data to make the local maxima appear periodic (Fig. \ref{fig:wavelength recalibration}).

\subsection{Fringe fitting}
\label{sec:fringe_fitting}

The model fitting used the package \verb|BayesicFitting| \citep{Kester2021}. The model used Bayesian inference to fit simultaneously the wavenumber correction, residual continuum variations, and two fringe components with finesse, frequency, and phase. 
In \verb|BayesicFitting|, models are compounded of simpler models.  In our case, we employed two \verb|EtalonModel| representing one fringe component each.  Both models had parameters scale, finesse, frequency, and phase. The ‘scale’ parameter of the first component was allowed to vary along wavelengths using a spline model to account for left-over continuum variations, and was set to unity for the second component. The finesse for fringe component 1 was allowed to vary smoothly with wavenumber using a spline model.  For fringe component 2, a single, constant finesse parameter was found sufficient to fit the data.  Wavenumber recalibration was initialized as described above and allowed to vary slightly during the fit; to this end, a spline model for wavenumber correction was piped into the compounded model describing the product of two fringe components.

In total, there were 22 model parameters: $f_1$, $\phi_1$, $f_2$, $F_2$, $\phi_2$, five coefficients for the spline model describing $F_1$, seven coefficients for the wavenumber recalibration, and five coefficients for the residual continuum.  All parameters were given a prior centred around values we deemed reasonable \citep[see for reference Tables 1 and 2 in][]{20ArRiRe} and appropriate constraints: finesse, for example, was constrained to be positive, while phases were circular to prevent the code from wasting time by sending a phase parameter a period up or down.

\section{Observations and data processing}
\label{sec:observations}

The MIRI FM ground test campaign, that took place in 2011 at the Rutherford Appleton Laboratory (RAL, UK), provided the first reference datasets to derive a representative set of fringe flats in the entire MRS wavelength range. These measurements were taken using the MIRI Telescope Simulator (MTS) tuneable blackbody calibration source. The MTS provided an illumination at the MIRI input with the same effective f-number and aperture mask shape as JWST. The tuneable blackbody calibration source produced a spatially extended illumination pattern at the MRS focal planes. The temperature of the blackbody could be tuned to values between $\sim$200~K and $\sim$800~K, temperatures of 800~K, 600~K, and 400~K were used for the tests, and for the reference fringe flats the 800~K measurements were used since these produced the highest $S/N$. In a first pass, the fringes were fitted using continuum normalisation and fitting a large number of sinusoids. This was later revised to use the Fabry-Pérot model described in Sect.~\ref{sect:math} and the methodology described in Sects.~\ref{sec:numerical model} and~\ref{sec:fringe_fitting}. After the FM campaign at RAL, MIRI was delivered to the NASA Goddard Space Flight Center, where three more cryo-vacuum (CV) test campaigns were completed. Although CV tests included high $S/N$ point source measurements, the Optical Ground Support Equipment (OGSE) did not include the possibility for extended source measurements with an extended source external to MIRI similar to the FM campaign. Due to the MRS internal calibration source illumination having a much smaller f-number compared to the JWST-incoming beam, the former results in a much more collimated beam on the detector and significantly larger fringes, as exemplified in Fig.~\ref{fig:washingOut}. For that reason, the FM datasets and FM-derived fringe flats remained the main reference until JWST-MIRI was in flight.

\begin{figure}[t!]
    \centering{\includegraphics[width=0.92\columnwidth]{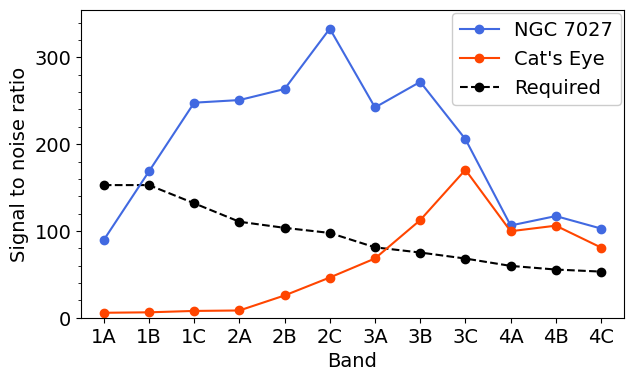}}
    \caption{Signal-to-noise ratio of individual pixels obtained from the coadded `\texttt{cal}'  images for NGC 7027 (blue) and Cat's Eye (red). The circles are the median $S/N$ over each band. The $S/N$ required to reach a fringe peak amplitude below 1\% on individual spaxels for extended sources after fringe flat calibration is shown in black; this required $S/N$ is that of the data that would be used to build the fringe flats.}
    \label{fig:SNR}
\end{figure}

The first batch of flight fringe flats were built from observations of the Cat's Eye Nebula (NGC~6543) that were obtained during commissioning (Program COM/MIRI 1031, PI: A. Labiano). These observations were made on 4 June 2022, they used the mosaic mode with four pointings, a four-point dither per sub-band, and one exposure per dither, resulting in 16 images per band. All pointings and dithers were coadded at the detector level (`\texttt{rate}' files) to maximise the $S/N$ before running the fringe fitting procedure. 
Fringe flats were derived for channels 3 and 4 (MIRIFULONG detector). The $S/N$ was too low for channels 1 and 2 (MIRIFUSHORT detector, see Fig. \ref{fig:SNR}), so fringe flats derived from FM-level ground test data simulating a spatially extended source were used instead. In addition, the Cat's Eye Nebula displays strong emission lines at numerous wavelengths, which posed difficulties to the fringe fitting because of scattered light inside the detector \citep[which produces the imager cruciform and large MRS PSF wings, as shown in Fig. 3 of][]{Argyriou2023} and the necessity to exclude certain wavelength ranges. These fringe flats were made available to science users immediately after the end of commissioning, when first science data became available (July 2022), and were in use on the JWST Calibration Reference Data System\footnote{\url{https://jwst-crds.stsci.edu}} (CRDS) until June 2023.

New calibration observations were obtained on the extended planetary nebula NGC~7027 (Jewel Bug Nebula) during Cycle~1 (Program CAL/MIRI 1523, PI: D. Law). The goal was to derive new fringe flats, and to obtain external flatfields (illumination flats) to provide low-spatial-frequency flatfields (L-flats) for the MIRI MRS in order to probe the overall system transmission and illumination across the detectors \citep[see discussion in Sect. 2 of][]{Law2025}.
The observations were made on 14 November 2022 and consisted of a $3\times3$ mosaic with 67\% overlap between adjacent pointings. For each pointing, a two-point dither with one exposure per dither was used for each sub-band. This resulted in 18 images per band. Each exposure contained ten integrations with five groups per integration, for a total science time of 7506 s (2.085 h) and a charged time of 20762 s (5.77 h). Background observations were also taken but were not used in this analysis.
NGC~7027 was not observable during the part of commissioning pertaining to MIRI spectrophotometric calibration so Cat's Eye was chosen instead.

We retrieved the level 1 output (`\texttt{rate}') files for NGC 7027 from the Mikulski Archive for Space Telescopes (MAST).  The CRDS version of this data was 11.16.14, the CRDS context was jwst\_1018.pmap, and the version of the JWST calibration pipeline that produced this level 1 output was 1.8.2 \citep{jwst2022}.  The data from the MIRIFUSHORT detector using the C (LONG) grating were further processed using the \verb|straylight| step in the level 2 pipeline to remove scattered light from saturated lines that contaminates neighbouring slices (this effect was the worst for band 2C). The version of the pipeline used to run the \verb|straylight| step was 1.9.4, the CRDS version was 11.16.20, and the CRDS context was jwst\_1063.pmap.  For each detector/band combination, all the output files were coadded to create a single high $S/N$ `\texttt{rate}' file. This produced six final `\texttt{rate}' files.  These files were the starting point of the fringe fitting procedure described in Sect. \ref{sec:numerical model}.  Care was taken to ignore wavelength ranges containing bright emission lines since the fringe-fitting code is not set up to account for them.  This was done in two passes: first based on quick-look extracted spectra (`\texttt{x1d}'  files) as downloaded from MAST, then based on reduced detector-level data.
This bright spatially extended source \citep[K = 7.305 mag, from WISE,][]{Cutri2013} provided a high enough $S/N$ to derive fringe flats for bands 1C to 4C. For bands 1A and 1B, the $S/N$ was too low (Fig. \ref{fig:SNR}) and the fringe flats did not improve compared to those obtained from FM ground test data, so these ground fringe flats were kept instead.

We computed the $S/N$ to assess the quality of extended source data to build fringe flats (these calculations do not provide the $S/N$ on the continuum of a final science spectrum, especially not for a point source or semi-extended source).
We used the surface brightness and error arrays provided in the \texttt{Spec2} pipeline products (SCI and ERR extensions of the `\texttt{cal}'  files), because they contain a more complete error budget than the `\texttt{rate}'  files. The individual `\texttt{cal}'  images were coadded into a single image in each band, with uncertainties computed as the quadratic sum of the uncertainties of individual images. We computed the $S/N$ as the ratio SCI/ERR of the coadded images and took the median in each band using only good pixels (with a data quality flag of 0). In other words, the $S/N$ was obtained from the surface brightness and its error on good pixels of the calibrated and coadded detector image for each band, that is, across the full field of view and along the full spectral range of that band. Only errors included in the \texttt{jwst} pipeline were taken into account. For these calculations, we retrieved the `\texttt{rate}'  files with \texttt{jwst} pipeline version 1.16.1, CRDS version 12.0.5, and CRDS context jwst\_1303.pmap from MAST, and built the `\texttt{cal}'  files with \texttt{jwst} pipeline version 1.17.1, CRDS version 12.0.9, and CRDS context jwst\_1322.pmap. The data used in this paper may be obtained from the MAST archive at
\url{http://dx.doi.org/10.17909/ha2f-1j70}

\section{Results}
\label{sec:results}

Best-fit fringe parameters were derived by fitting the numerical model described in Sect.\ \ref{sec:model} to the data described in Sect.\ \ref{sec:observations}. These fringe parameters were then used to build the fringe flats. The fringe flat in each band was normalised to have a maximum equal to unity, and they were bound by pairs to create six $1032\times1024$ pixel images that correspond to the two detectors and three bands.
NGC 7027 was much more suited to build fringe flats than Cat's Eye because it is much brighter: in 1C the typical continuum count rates are about 0.4 DN/s for Cat's Eye and 60 DN/s for NGC 7027, and in 4B they are 14 DN/s for Cat's Eye and 310 DN/s for NGC 7027.
The $S/N$ of the coadded `\texttt{cal}'  images is shown in Fig. \ref{fig:SNR} together with the $S/N$ required to build fringe flats (Sect.~\ref{sec:signal-to-noise ratio}). 
The current MRS fringe flats available on CRDS (jwst\_miri\_fringe\_0070 to jwst\_miri\_fringe\_0075) were built from NGC~7027 for bands 1C to 4C and from FM ground test data for bands 1A and 1B. They were released in June 2023. Since then, fringe flats derived from the Cat's Eye nebula are no longer in use.
In the JWST calibration pipeline, the fringe flat calibration is applied by the \texttt{fringe} step of the \texttt{Spec2Pipeline} and consists of dividing the science data by the fringe flat for each detector and band. This calibration occurs at the detector level, that is on the 2D detector images.
The \texttt{fringe} step occurs before the photometric calibration step (\texttt{photom}), so the pixel values obtained after division by the static fringe flats are still in count rate units (DN/s). It can also be noted that the \texttt{residual\_fringe} step occurs after the \texttt{photom} step (and so does the optional 1D residual fringe correction in the \texttt{Spec3Pipeline}), so these provide physical surface brightnesses.

\begin{figure}
    \includegraphics[width=0.99\columnwidth]{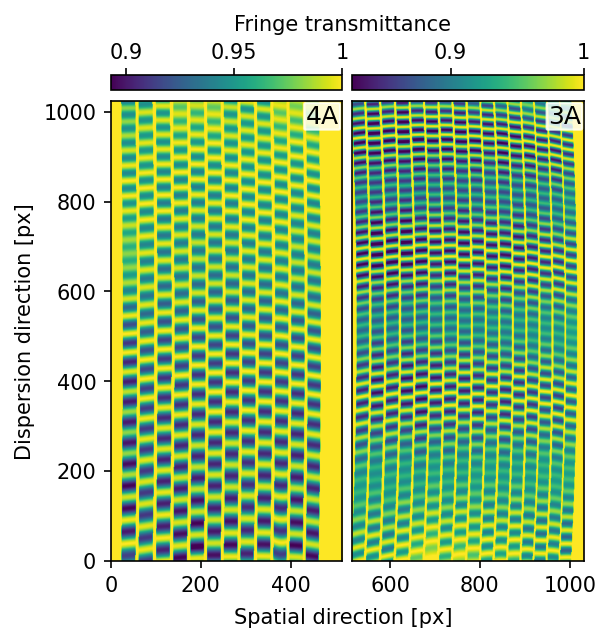}
    \caption{Fringe flat for detector MIRIFULONG band A derived from observations of NGC~7027. Bands 4A and 3A are on the left and right hand side, respectively. The colour scale for each half-detector is indicated at the top. The two half-detectors are split for clarity. The beating is apparent for band 3A.}
    \label{fig:fringe-flat}
\end{figure}

\begin{figure}
    \includegraphics[width=\columnwidth]{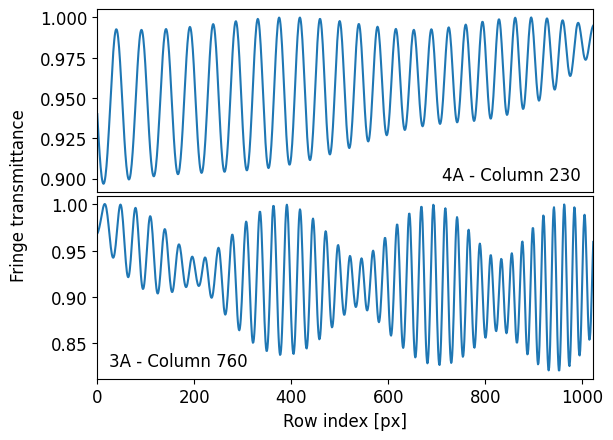}
    \caption{Vertical cuts along columns 230 (band 4A, top) and 760 (band 3A, bottom) of the fringe flat shown in Fig. \ref{fig:fringe-flat}. One fringe component dominates for band 4A whereas the two components produce beating for band 3A.}
    \label{fig:columnCut-flat}
\end{figure}

\begin{figure}[h!]
    \includegraphics[width=\columnwidth]{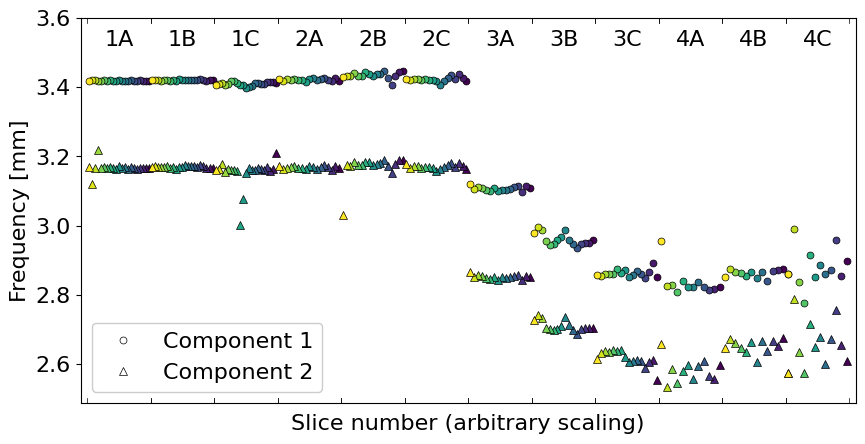}
    \caption{Frequency of the first (circles) and second (triangles) fringe component for all channels and bands. Colours represent individual slices, ordered by increasing slice number from purple to yellow. Bands 1A and 1B are from FM ground test data, bands 1C to 4C are from NGC 7027 data.}
    \label{fig:period-allbands}
\end{figure}

\begin{figure}[h!]
    \includegraphics[width=\columnwidth]{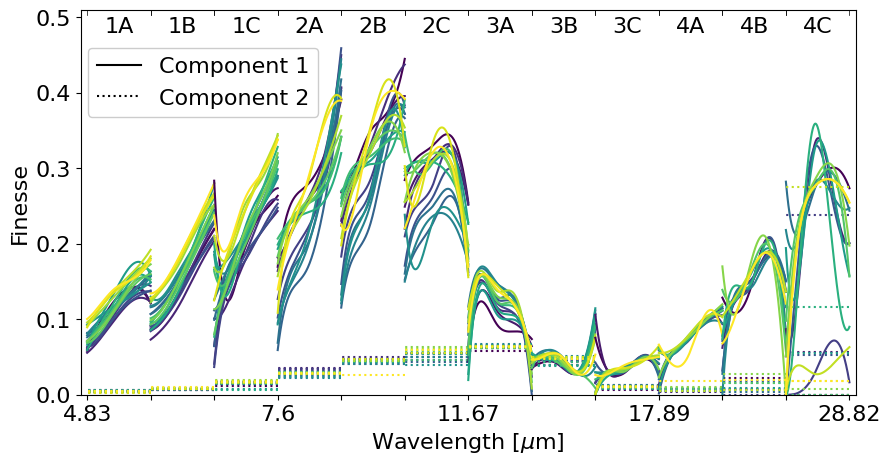}
    \caption{Finesse as a function of wavelength for all channels and bands. The finesse of the first fringe component is modelled by a five coefficient spline along each slice (plain lines) and that of the second fringe component is a constant (dotted lines). Colours represent individual slices, ordered by increasing slice number from purple to yellow. Bands 1A and 1B are from FM ground test data, bands 1C to 4C are from NGC 7027 data. For band 3B, the finesses of the two components have similar values, creating the strongest beating.}
    \label{fig:finesse-allbands}
\end{figure}

The fringe flat obtained for bands 3A and 4A (MIRIFULONG detector, SHORT band) are displayed in Fig. \ref{fig:fringe-flat} and examples of cuts along columns are shown in Fig. \ref{fig:columnCut-flat}. The correction of science data by the fringe flat is illustrated in Figs. \ref{fig:rawFringes} and \ref{fig:columnCut-fringecor}, where a science 2D detector image and cuts along columns are shown before and after correction by the corresponding fringe flat.
The derived frequency and finesse of both fringe components are shown for all channels and bands in Figs. \ref{fig:period-allbands} and \ref{fig:finesse-allbands}.

Focusing on Fig. \ref{fig:period-allbands}, in bands 1A to 2C (MIRIFUSHORT detector), the frequency of Component 1 probes the optical thickness between the AR coating and the pixel metalization, and the frequency of Component 2 probes the optical thickness of the cavity between the AR coating and the buried contact. The difference between the two frequencies probes the optical thickness of the cavity between the buried contact and the pixel metalization. The two components do not match directly the two cavities: fringes from the buried contact to pixel metalization cavity are not resolved, but they beat against fringes from the AR coating to buried contact cavity, and that beating is measured. 
Bands 3A to 4C are measured using a different detector (MIRIFULONG) with different geometric and optical properties. Band 3A to 3C are a transitionary stage where the buried contact becomes significantly reflective \citep[see Fig. 6 and 7 in][]{20ArRiRe} and the AR coating becomes most efficient (least reflective) at 16 $\mu$m. In bands 4A to 4C, the detector infrared-active layer is at its most absorptive, and in fact Component 1 is now the cavity between the AR coating and the buried contact, which takes over due to a minimal amount of photons that survive after they enter the cavity between the buried contact and the pixel metalization. Because of this, the fit on the frequency of Component 2 becomes very noisy.

Studying Fig. \ref{fig:finesse-allbands}, in bands 1A to 2C, the finesse of Component 1 increases linearly due to the increase in the reflectivity of the AR coating. The jumps in the finesse between the bands are caused by the change in the spectral resolution (lower spectral resolution results in lower fringe amplitude). The finesse of Component 2 also increases due to the linearly increasing reflectivity of the buried contact. In bands 3A to 4C, the finesse profile of Component 1 is ruled by the reflectivity of the AR coating, optimized to be minimal at 16 $\mu$m. The low quantum yield of the MRS detectors in the longer wavelengths of band 4C results in low recorded signals, larger uncertainty in the fits, and that affects the polynomial fit across band 4C \citep[see][for a discussion about the quantum yield of Si:As blocked impurity band detectors]{Woods2011}.

\section{Efficiency of the fringe correction}
\label{sec:efficiency}

\subsection{Method}

To measure the efficiency of the fringe correction based on the NGC 7027 data (or FM ground test data for bands 1A and 1B), we evaluated the fringe amplitude before and after applying the fringe flats. We used data of extended, semi-extended, and point sources as described in the next sections. For each object, we started from the `\texttt{rate}' files and ran the \texttt{jwst} \texttt{Spec2Pipeline} with or without the fringe flat step (\texttt{fringe}). We skipped the background subtraction and the steps that come after the fringe flat step, namely the spectrophotometric calibration and residual fringe correction. Then, we ran the \texttt{jwst} \texttt{Spec3Pipeline} skipping all the steps but the \texttt{cube\_build} step to build one cube per band. Finally, we did an additional run with the 2D residual fringe correction step turned on in the \texttt{Spec2Pipeline} to also evaluate its effect. The 1D residual fringe correction was not used.  We used the coadded images for extended sources and only one dither for point and semi-extended sources (because for these the fringe pattern differs between dithers so coadding them would alter the fringes). Each band was analysed independently. 
Fringes were measured on individual spaxels of the reconstructed 3D cubes and on individual columns (dispersion direction) of the 2D detector images, where the interference occur. Each of them is a spectrum. In the 3D cubes, we used 16 spaxels evenly distributed over the field of view for extended sources and the brightest 16 spaxels for semi-extended and point sources. For extended sources, the normalised fluxes as a function of wavenumber of each column within a slice of a 2D detector image are very similar (the fringe pattern is nearly the same) and the fringes are also similar between slices (except for the phase that can shift). For point sources, the signal from the source is contained in only a few slices and in a small number of columns within those slices. In addition, because the slices are bent on the detector, the flux can vary steeply along a given column and disentangling the fringes from the continuum is challenging (particularly for channels 2 and 3 where the bending is the worst out of the four channels). 
On individual spaxels, this slice bending yields a sampling effect that manifests as large flux variations particularly at the shortest and longest wavelengths of the band, where the bending is the strongest.
The process of averaging the signal of multiple pixels in the detector image into one spaxel slightly affects the fringe properties (the cleanest fringe measurement comes from the detector image). However, because of the slice bending, working on spaxels was more convenient for point sources (and as convenient as working on columns for extended sources). In the following, we show the results obtained on spaxels. 
For point and semi-extended sources, fringes were measured only in a central wavelength region of each band that was not affected by the sampling effect.

To measure fringes, we resampled the spectrum into a regular grid of 40 points per wavenumber by taking the median flux in each bin (this spectrum is referred to as the `resampled spectrum' in the following) and searched for local maxima and minima using the same procedure as in Sect. \ref{sec:numerical model} (Figs. \ref{fig:flux-wtops-3B-slice3} and \ref{fig:flux-wtops-4B-slice10}). These `tops' correspond to the envelope of the oscillations caused by fringes; an example for NGC 7027 is shown in Fig. \ref{fig:tops-allbands-2types-NGC-7027}. Then, we gathered the tops identified on all 16 spaxels, subtracted a value of one to centre them on zero, took the absolute values to flip the negative tops with respect to zero (the positive tops being unchanged), and computed the 16th, 50th, and 84th percentiles of that distribution as an estimate of the fringe amplitude and its variations. Here and in the following Tables and Figures, the amplitude is defined as the peak deviation from the mean (that is half of the peak-to-peak amplitude). As shown in Fig. \ref{fig:tops-allbands-2types-NGC-7027}, the fringe amplitude varies along wavelength within a band either monotonically or with beating, thus the 16th -- 84th percentile interval should be considered as a range rather than as an uncertainty. 
For the percentiles of band 4C, the longest wavelengths ($> 27.5 \,\mu$m, one quarter of the band) were not used because the signal becomes too low to measure fringes, for all sources used here.

To assess the results, we also calculated the standard deviation of the full spectrum and that of the resampled spectrum.
We also measured the root mean square (RMS) of the tops after subtracting the resampled spectrum mean (that is nearly one) and divide it by$\sqrt{2}$. This gives an estimate of the spectrum standard deviation but is less sensitive to high frequency variations, random noise, or weak underlying forests of lines.
In addition, we computed the periodogram of the resampled spectrum using the Welch's method \citep{Welch1967} as implemented in \texttt{scipy}. We chose the `spectrum' scaling that computes the squared magnitude spectrum and a `flattop' window: with these settings, the square root of the peak height gives an estimate of the RMS amplitude of the signal at each sampled frequency, which gives another measurement of the fringe amplitude. The frequency of the main peak was compared to the known frequencies of the fringes (Fig. \ref{fig:period-allbands}) to check if the dominant signal was due to fringes or to other variations. 
The Welch's method divides the input signal into segments, computes a periodogram for each segment, and averages the periodograms. We set the length of each segment to $n / N$ where $n$ is the number of points in the spectrum and $N$ is 1, 2, or 3. We set the overlap between segments to half their length (the default). The main peak amplitude differs depending on $N$ and only accounts for the power at that frequency, thus we used this approach only as a comparison but not as a main metric.
These RMS were computed for individual spaxels, then they were averaged over the 16 spaxels. 

Although dithers are averaged and spaxels are summed when extracting science spectra, the goal here was to measure the efficiency of the fringe correction, not the way they average in final spectra. 
Thus, we did not average nor sum the spectra obtained on individual spaxels before measuring fringes because that would average the fringe patterns in an uncontrolled way. 
It should be noted that the dichroic fringes have high frequencies and are largely smoothed out in the resampled spectrum, so the calculations performed on the resampled spectrum do not account for dichroic fringes.

\subsection{Application to extended sources}

\subsubsection{Application to NGC 7027}

\begin{figure}
    \includegraphics[width=\columnwidth]{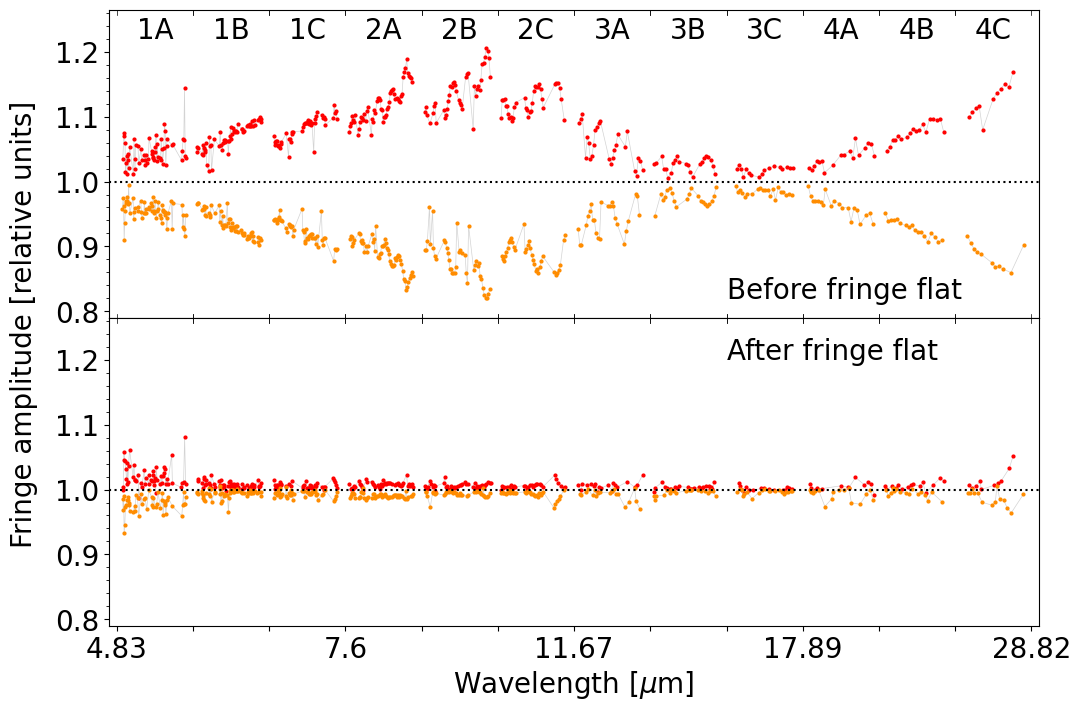}
    \caption{Local extrema of a spectrum of NGC 7027 before (top) and after (bottom) fringe flat correction measured on a single spaxel in each band. The red and orange points correspond to the upper and lower envelopes of the fringes, respectively. Both panels are on the same scale. Fringes are weaker in 1A-1B and 3B-3C because the AR coating is most efficient at 6 and $16\,\mu$m. A few peaks are present and are due to spectral lines.}   
    \label{fig:tops-allbands-2types-NGC-7027}
\end{figure}

\begin{figure}
    \includegraphics[width=\columnwidth]{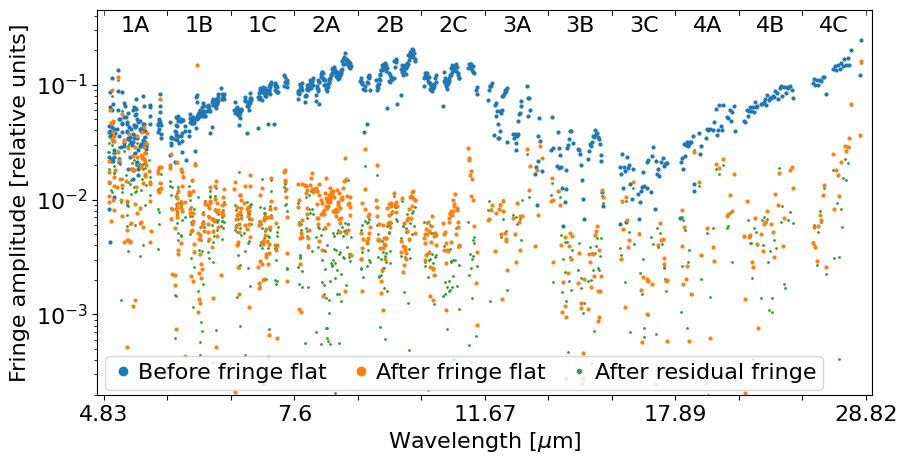}
    \caption{Local extrema of a spectrum of NGC 7027 before (blue) and after (orange) fringe flat correction and after 2D residual fringe correction (green) measured on a single spaxel in each band. The continuum-corrected normalised spectra have been shifted to a zero mean and the local minima have been flipped to their absolute value.}
    \label{fig:tops-allbands-NGC-7027}
\end{figure}

\begin{figure}
    \centering
    \includegraphics[width=0.95\linewidth]{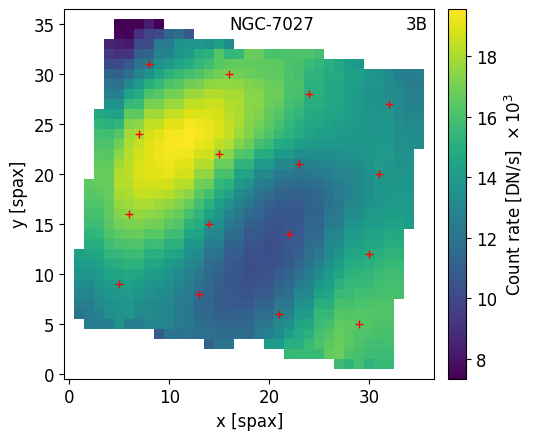}
    \caption{Median of the coadded image of NGC 7027 for band 3B (median of the flux cube over wavelengths). The spaxels used to measure fringes in that band are shown as red crosses. The colour scale is indicated on the right (individual images would have approximately 1/18th of this flux).}
    \label{fig:spax-use-3B-NGC-7027}
\end{figure}

First, the fringe flat correction efficiency was measured on NGC 7027. These observations were used to build the fringe flats except for bands 1A and 1B. This provides a check of our method and of the quality of the fringe flats. 
For the extended sources NGC 7027 and Cat's Eye, we discarded wavelength regions that contain spectral features as was done in the fringe flat building process.
The tops identified for an example spaxel in each band before and after the fringe flat step are shown in Figs. \ref{fig:tops-allbands-2types-NGC-7027} and \ref{fig:tops-allbands-NGC-7027}, the latter also shows the tops after the 2D residual fringe correction step. The median image (median of the flux cube over wavelengths) and the spaxels that were used for band 3B are shown in Fig. \ref{fig:spax-use-3B-NGC-7027} as an example. The median fringe amplitude and its range are reported in Table \ref{tab:fringe correction NGC-7027}. 

Before fringe flat calibration, the median fringe amplitude varies from about 2 to 12\% depending on the band, with significant variations in each band. After fringe flat calibration, the median amplitude is below 1\% and the 84th percentile is below 2\% in all bands except 1A and 4C. The correction is the strongest in channel 2 where the fringes are reduced by a factor 15 to 25, and the amplitude is the lowest (0.3\%) in bands 3B and 3C (these bands already had the lowest amplitude before fringe flat calibration). For band 1A, the periodogram after fringe flat indicates that the power is spread over many frequencies and not specifically at those of fringes because of the high noise level. This band would clearly benefit from a flight fringe flat obtained from a source of higher $S/N$ compared to ground tests. The residual fringe correction reduces fringes mostly in channel 2 and bands 4B $-$ 4C, where fringes are still visible after the fringe flat calibration, and to less extent in 1B $-$ 1C. It is less or not efficient when the spectrum is noisy (1A) or when the continuum has non-sinusoidal variations on wavelength scales similar to or a few times larger than fringes (4A). These variations were not well removed by our continuum removal procedure (that is designed to keep the fringes and remove long period, smooth variations) nor by the residual fringe correction (that removes only sinusoids).

\begin{figure}
    \centering
    \includegraphics[width=\linewidth]{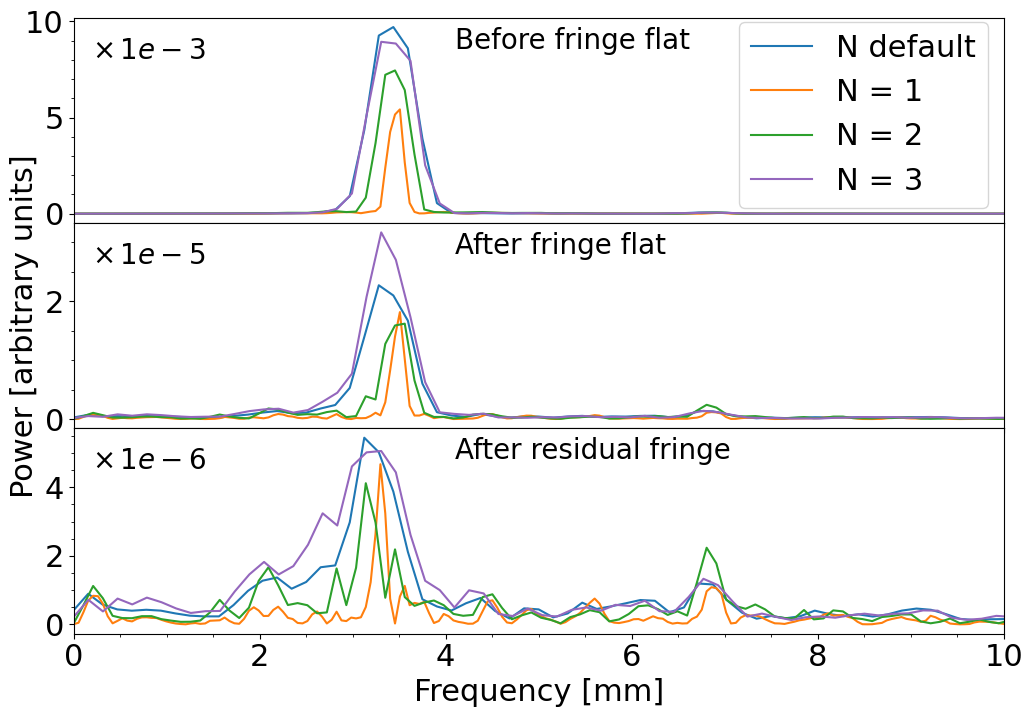}
    \caption{Periodograms of a spectrum of NGC 7027 before (top) and after (middle) fringe flat correction and after residual fringe correction (bottom) obtained from a single spaxel of band 2A. The frequency of the main peak matches that derived for the fringes and its amplitude decreases after applying the fringe corrections, which indicates that the fringes have been reduced. The region with frequencies between 10 and 20 mm is nearly flat and is not shown.}
    \label{fig:periodograms-NGC-7027}
\end{figure}

\begin{figure}
    \centering
    \includegraphics[width=1\linewidth]{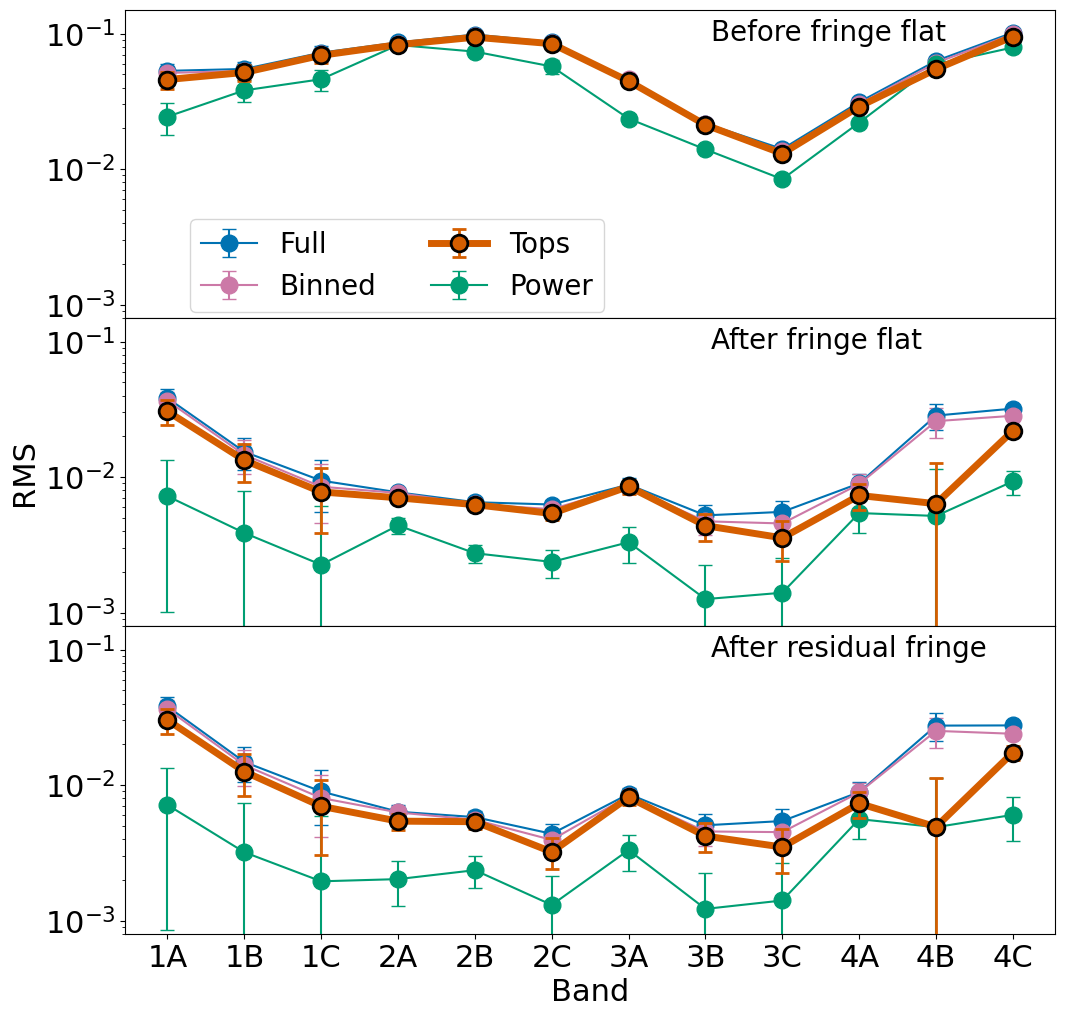}
    \caption{RMS for NGC 7027 for all bands before (top) and after (middle) fringe flat correction and after residual fringe correction (bottom) averaged over the 16 spaxels. The various metrics are the standard deviation of the full spectrum (blue), the standard deviation of the binned spectrum (pink), the tops RMS (orange), and the RMS derived from the main peak of the periodograms (green). The error bars indicates the dispersion of these quantities among spaxels and, for the periodograms, among the four values of $N$.} 
    \label{fig:rms-NGC-7027}
\end{figure}

Example periodograms are shown in Fig. \ref{fig:periodograms-NGC-7027} for band 2A. The frequency and amplitude accuracy is limited by the sampling and the choice of $N$. Nevertheless, it can be seen that the frequency of the main peak closely matches that derived for the fringes (Fig. \ref{fig:period-allbands}) and its amplitude decreases after applying the fringe corrections. For that band, fringes are largely reduced but are still dominant even after residual fringe correction. 
The various RMS measurements are shown in Fig. \ref{fig:rms-NGC-7027}. The RMS obtained from the full and resampled spectra and from the tops are similar (except in band 4B after fringe correction). 
The RMS obtained from the periodograms was averaged over the four values of $N$, which results in large uncertainties. It follows the same trends as the other RMS values but it is lower. In particular, after fringe flat, a larger fraction of the power is at frequencies other than that of the main peak. It may indicate that the residual variations are not strongly dominated by one fringe component, or by the fringes studied here, or they are only dominant in small regions of the spectrum, or that other noise sources become relatively more important as fringes decrease. Visual inspection of band 3C indeed shows that these fringes are generally very well suppressed (they are still visible at the longest wavelengths in a few spaxels), whereas the high frequency fringes from the dichroic filters become important. In contrast, for band 2A, the residual variations are still due to the fringes studied here and further decreases after residual fringe correction.

\begin{table*}
\begin{center}
\caption{Effect of the fringe correction for NGC 7027.}
\label{tab:fringe correction NGC-7027}
\small
\begin{tabular}{c|c|c|c|c}
\hline\hline
\multirow{2}{*}{Channel} & \multirow{2}{*}{Band} & \multicolumn{3}{c}{Amplitude (\%)} \\
\cline{3-5}
     & & No correction & Fringe flat & Residual fringe \\
\hline
\multirow{3}{*}{1} & A & $4.8 \;\; [3.1 - 6.69]$ & $1.96 \;\; [0.93 - 3.88]$ & $2.03 \;\; [0.93 - 3.83]$ \\ 
                    & B & $6.56 \;\; [4.42 - 8.84]$ & $0.91 \;\; [0.42 - 1.7]$ & $0.77 \;\; [0.35 - 1.49]$ \\ 
                    & C & $8.86 \;\; [6.38 - 11.56]$ & $0.66 \;\; [0.3 - 1.18]$ & $0.58 \;\; [0.26 - 1.05]$ \\ 
\hline
\multirow{3}{*}{2} & A & $10.78 \;\; [8.72 - 13.42]$ & $0.72 \;\; [0.38 - 1.14]$ & $0.45 \;\; [0.19 - 0.85]$ \\ 
                    & B & $12.4 \;\; [9.84 - 15.32]$ & $0.54 \;\; [0.28 - 0.98]$ & $0.46 \;\; [0.22 - 0.83]$ \\ 
                    & C & $11.35 \;\; [8.96 - 13.65]$ & $0.47 \;\; [0.21 - 0.88]$ & $0.31 \;\; [0.14 - 0.56]$ \\ 
\hline
\multirow{3}{*}{3} & A & $5.45 \;\; [2.77 - 8.61]$ & $0.69 \;\; [0.31 - 1.44]$ & $0.71 \;\; [0.32 - 1.3]$ \\ 
                    & B & $2.81 \;\; [1.39 - 3.6]$ & $0.32 \;\; [0.11 - 0.83]$ & $0.3 \;\; [0.14 - 0.73]$ \\ 
                    & C & $1.75 \;\; [1.13 - 2.2]$ & $0.29 \;\; [0.11 - 0.64]$ & $0.3 \;\; [0.11 - 0.63]$ \\ 
\hline
\multirow{3}{*}{4} & A & $3.66 \;\; [2.33 - 5.41]$ & $0.53 \;\; [0.16 - 1.86]$ & $0.56 \;\; [0.14 - 1.74]$ \\ 
                    & B & $7.4 \;\; [6.01 - 9.2]$ & $0.63 \;\; [0.22 - 1.27]$ & $0.48 \;\; [0.18 - 0.85]$ \\ 
                    & C & $12.03 \;\; [10.22 - 14.62]$ & $1.13 \;\; [0.48 - 2.48]$ & $0.59 \;\; [0.22 - 1.29]$ \\ 
\hline
\hline
\end{tabular}
\normalsize
\end{center}
\textbf{Notes}. For each MRS channel and band, the median of the fringe peak amplitude measured from the tops is given and its 16th and 84th percentiles are shown in brackets. Values are shown before correction, after calibration by the fringe flat, and after residual fringe correction. All numbers are in percent.
\end{table*}

\subsubsection{Application to Cat's Eye}

Here we used the Cat's Eye nebula observations calibrated with the fringe flats derived from NGC 7027. The analysis was restricted to channels 3 and 4 to have enough signal to analyse fringes (Fig. \ref{fig:SNR}). The results are displayed in Table \ref{tab:fringe correction Cats-Eye}. The fringes are reduced by a factor three to ten depending on the band. The amplitudes are larger than for NGC 7027 but it is due to the lower data quality rather than to a less efficient fringe correction: the continuum has more variations, spectral lines are numerous, and the flux is lower than for NGC 7027. After fringe flat calibration, in band 3A, the spectrum is dominated by noise (not fringes). In bands 3B and 3C, the fringes are well corrected, residual fringes are visible but are mixed with noise (3B) or with fringes from the dichroics (3C). In 4A, the continuum has variations that are not well removed, 4B is affected by a few spectral lines, and in 4C only a small part of the spectrum is used. Overall, the fringe flat calibration is also efficient on Cat's Eye but the fringe measurements are affected by the lower data quality.

\begin{table*}
\begin{center}
\caption{Effect of the fringe correction for Cat's Eye, restricted to channels 3 and 4 where the source has enough signal.}
\label{tab:fringe correction Cats-Eye}
\small
\begin{tabular}{c|c|c|c|c}
\hline\hline
\multirow{2}{*}{Channel} & \multirow{2}{*}{Band} & \multicolumn{3}{c}{Amplitude (\%)} \\
\cline{3-5}
     & & No correction & Fringe flat & Residual fringe \\
\hline
\multirow{3}{*}{3} & A & $5.83 \;\; [3.21 - 9.54]$ & $2.05 \;\; [0.75 - 4.96]$ & $2.0 \;\; [0.77 - 4.71]$ \\ 
                    & B & $3.02 \;\; [1.59 - 4.28]$ & $0.77 \;\; [0.28 - 1.61]$ & $0.77 \;\; [0.31 - 1.62]$ \\ 
                    & C & $1.5 \;\; [0.93 - 2.3]$ & $0.43 \;\; [0.15 - 0.94]$ & $0.43 \;\; [0.16 - 0.89]$ \\ 
\hline
\multirow{3}{*}{4} & A & $4.19 \;\; [2.93 - 6.06]$ & $0.99 \;\; [0.22 - 2.34]$ & $0.99 \;\; [0.21 - 2.32]$ \\ 
                    & B & $7.36 \;\; [5.12 - 9.77]$ & $0.79 \;\; [0.32 - 2.66]$ & $0.76 \;\; [0.27 - 1.85]$ \\ 
                    & C & $13.67 \;\; [12.28 - 15.28]$ & $2.0 \;\; [1.06 - 2.93]$ & $0.91 \;\; [0.31 - 1.72]$ \\ 
\hline
\hline
\end{tabular}
\normalsize
\end{center}
\textbf{Notes}. See Table \ref{tab:fringe correction NGC-7027} for explanations.
\end{table*}

\subsection{Application to point sources}

\begin{figure}
    \centering
    \includegraphics[width=0.95\linewidth]{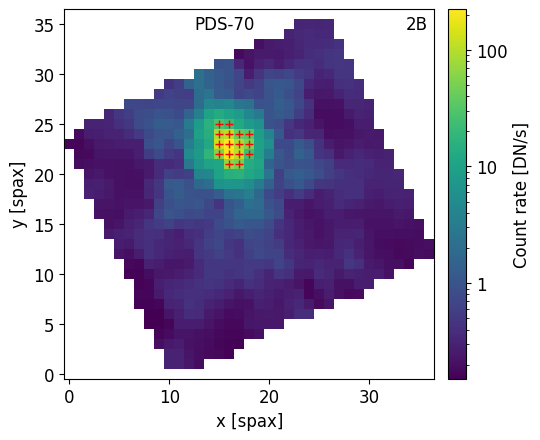}
    \caption{Median image of PDS~70 for band 2B (median of the flux cube over wavelengths). The spaxels used to measure fringes in that band are shown as red crosses. The colour scale is indicated on the right.}
    \label{fig:spax-use-3B-PDS-70}
\end{figure}

Fringe correction methods for point sources are presented in \citet{Argyriou2020} and \citet{Gasman2023a}. Extended source fringe flats provide only partial correction of fringes for point sources because the fringe patterns differ from those of extended sources (Sect. \ref{sec:Point sources versus extended sources}). We evaluated the fringe flat calibration on a point source using observations of PDS~70 that were taken on 1 August 2022 as part of the MIRI mid-INfrared Disk Survey (MINDS) Guaranteed Time Observations (GTO) program \citep[PID 1282, PI: Th. Henning;][]{Kamp2023, Henning2024}. An example median image and the 16 brightest spaxels are shown in Fig. \ref{fig:spax-use-3B-PDS-70}.
The fringe amplitudes and ranges are reported in Table \ref{tab:fringe correction PDS-70}.
As already noted in \citet{Argyriou2020} and \citet{Gasman2023a}, the fringe amplitude and phase depend on what part of the PSF is sampled by the pixel, they differ between adjacent columns in the detector image, and thus between spaxels. The amplitude can vary by a factor of two or more, particularly in channels 1 and the short wavelength end of channel 2 where the PSF is undersampled by design. Some columns within a slice can have fringes of lower amplitudes than the fringe flat while others can have fringes of larger amplitudes. The fringe flat has amplitudes that are roughly similar between columns, thus it cannot provide a good correction for all columns and can even change the fringe pattern (for example it can flip the phase if the amplitude in the fringe flat is larger than a certain value compared to the column to which it is applied). This may help fringes average out when summing spaxels, but it is not a reliable correction.

\begin{figure}
    \centering
    \includegraphics[width=\linewidth]{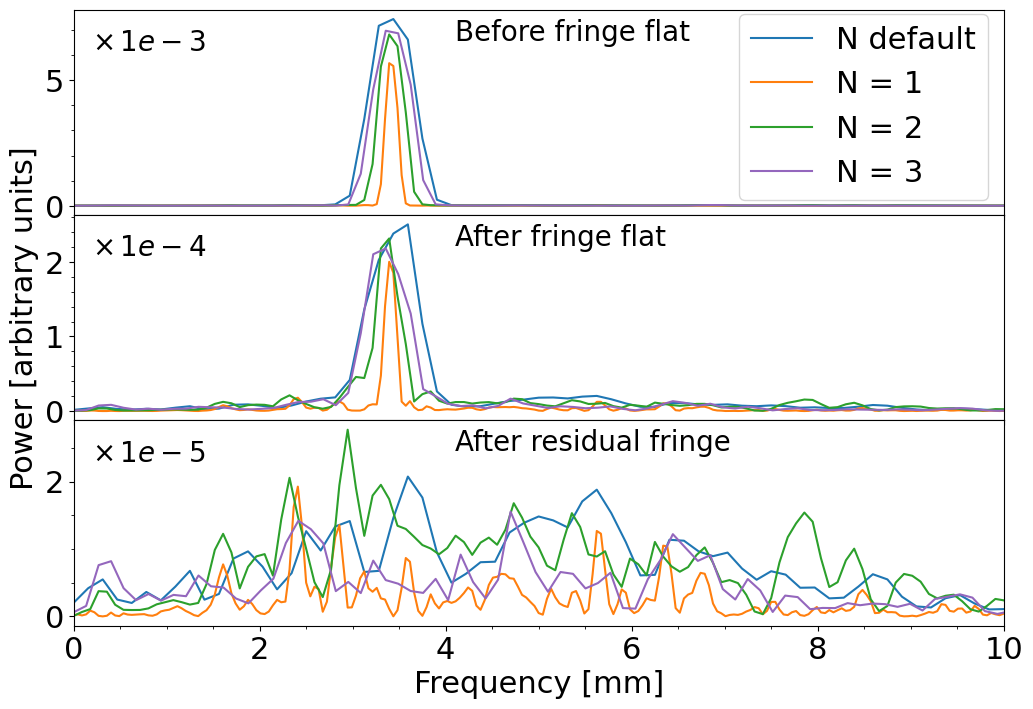}
    \caption{Same as Fig. \ref{fig:rms-NGC-7027} for PDS~70, for a single spaxel of band 1C. The fringe flat and residual fringe correction reduce the amplitude of the main peak, and the variations are no longer sinusoidal (in wavenumber space) after the residual fringe correction has been applied.}
    \label{fig:periodograms-PDS-70}
\end{figure}

As expected, the fringe flat calibration is less efficient than for extended sources. The fringe flat reduces the fringe amplitude by a factor two to five depending on the band. 
The correction is limited when the fringe amplitude varies the most between columns, in particular at short wavelengths (channels 1 and 2) and to a less extent in channel 4.
In band 3A, the initial fringe amplitude appears smaller than for extended sources, this is because the spectral region used to measure fringes is where the beating envelope is the narrowest. After fringe flat, the amplitude is slightly overestimated because the continuum is not completely removed.
The fringe flat performs best in bands 3B and 3C. In this channel, the central region not affected by the sampling effect is the smallest (one fifth of the band) and the continuum is not fully removed, so the actual fringe amplitudes are smaller than reported and visual inspection shows that fringes are well suppressed. For all channels and bands, visual inspection of the spectra shows that the fringe correction in regions that are affected by the sampling effect is similar to that of the central region. In channel 4, the background becomes significant and contributes to the fringe pattern.
When fringes remain, the residual fringe correction reduces them to median values between 0.7 and 2\%. It is most efficient in channels 1 and 2 where the initial fringe amplitude varies the most between columns (the 2D residual fringe correction is computed column-by-column).
Example periodograms for band 1C are shown in Fig. \ref{fig:periodograms-PDS-70}. The main peak amplitude decreases after applying the fringe flat and residual fringe correction. After residual fringe correction, the power becomes spread over many frequencies because the variations that remain are no longer sinusoidal.
As for extended sources, the dichroic fringes become a significant limitation in channels 3 and 4. Their amplitude is larger than for extended sources and they remain after the fringe flat and residual fringe correction.

As a comparison, we also analysed the observations of asteroid 526 Jena that were taken on 21 September 2023 (PID 1549, PI: K. Pontoppidan). The results are very similar to PDS~70 (except in bands 1A and 1B where the asteroid is faint).

\begin{table*}
\begin{center}
\caption{Effect of the fringe correction for PDS~70.}
\label{tab:fringe correction PDS-70}
\small
\begin{tabular}{c|c|c|c|c}
\hline\hline
\multirow{2}{*}{Channel} & \multirow{2}{*}{Band} & \multicolumn{3}{c}{Amplitude (\%)} \\
\cline{3-5}
     & & No correction & Fringe flat & Residual fringe \\
\hline
\multirow{3}{*}{1} & A & $6.13 \;\; [4.29 - 8.9]$ & $3.39 \;\; [1.59 - 7.49]$ & $1.44 \;\; [0.55 - 3.68]$ \\ 
                    & B & $7.26 \;\; [4.47 - 11.77]$ & $4.27 \;\; [2.2 - 7.46]$ & $1.54 \;\; [0.5 - 3.73]$ \\ 
                    & C & $10.5 \;\; [7.58 - 15.25]$ & $5.87 \;\; [2.73 - 9.78]$ & $1.46 \;\; [0.57 - 3.21]$ \\ 
\hline
\multirow{3}{*}{2} & A & $10.72 \;\; [7.51 - 15.54]$ & $4.07 \;\; [2.01 - 7.69]$ & $1.82 \;\; [0.73 - 4.3]$ \\ 
                    & B & $12.28 \;\; [9.09 - 15.42]$ & $3.07 \;\; [1.75 - 5.26]$ & $1.14 \;\; [0.37 - 2.93]$ \\ 
                    & C & $12.11 \;\; [9.38 - 14.66]$ & $2.39 \;\; [0.93 - 4.3]$ & $0.77 \;\; [0.28 - 2.05]$ \\ 
\hline
\multirow{3}{*}{3} & A & $3.6 \;\; [1.89 - 6.68]$ & $1.77 \;\; [0.51 - 4.31]$ & $1.56 \;\; [0.56 - 4.22]$ \\ 
                    & B & $3.2 \;\; [1.72 - 4.73]$ & $1.11 \;\; [0.41 - 2.51]$ & $1.08 \;\; [0.4 - 2.4]$ \\ 
                    & C & $1.67 \;\; [0.89 - 2.76]$ & $0.8 \;\; [0.31 - 2.04]$ & $0.73 \;\; [0.29 - 1.85]$ \\ 
\hline
\multirow{3}{*}{4} & A & $3.96 \;\; [2.39 - 6.42]$ & $1.65 \;\; [0.7 - 4.07]$ & $1.46 \;\; [0.53 - 3.52]$ \\ 
                    & B & $8.0 \;\; [5.78 - 10.47]$ & $2.43 \;\; [0.98 - 4.79]$ & $1.38 \;\; [0.57 - 3.37]$ \\ 
                    & C & $12.13 \;\; [9.85 - 15.27]$ & $2.74 \;\; [1.33 - 4.66]$ & $1.65 \;\; [0.62 - 3.36]$ \\ 
\hline
\hline
\end{tabular}
\normalsize
\end{center}
\textbf{Notes}. See Table \ref{tab:fringe correction NGC-7027} for explanations.
\end{table*}

\subsection{Application to semi-extended sources}

\begin{figure}
    \centering
    \includegraphics[width=0.95\linewidth]{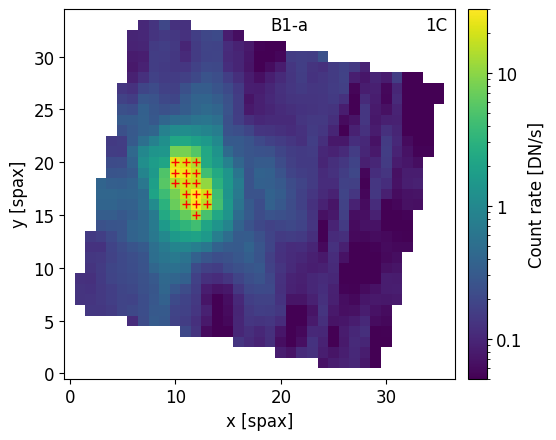}
    \caption{Median image of B1-a for band 1C (median of the flux cube over wavelengths). The spaxels used to measure fringes in that band are shown as red crosses. The colour scale is indicated on the right.}
    \label{fig:spax-use-1C-B1a}
\end{figure}

Semi-extended sources are the most problematic for fringe fitting. Fringe correction methods for extended sources and for point sources are different. Applying fringe flats to point sources reduces the fringes only partially, and the fringe flats derived for point sources \citep{Gasman2023a} are not suited to extended sources because the illumination is different and therefore the fringe pattern is different. We evaluated the fringe correction on \mbox{[HKM99] B1-a} that was observed on 12 September 2023 as part of the JWST Observations of Young protoStars (JOYS) GTO program (PID 1290, PI: E. van Dishoeck). This object is composed of two protostars and the observations are described in \citet{vanGelder2024}. The presence of two bright objects and potential structures in the protoplanetary disks lead to spatial inhomogeneities in the field of view. An example median image and the 16 brightest spaxels are shown in Fig. \ref{fig:spax-use-1C-B1a}. At short wavelengths, the source is seen as two bright regions, whereas at longer wavelengths it is similar to a single point source with diffuse emission on the sides.
The fringe amplitudes and ranges are reported in Table \ref{tab:fringe correction B1-a}.
The fringe corrections are less efficient than for extended objects. The central stars dominate the overall flux, and the fringe patterns and correction efficiencies are similar to those of a single point source. Fringes were measured on the brightest spaxels; instead one could study spaxels in less bright regions, for example in an annulus that excludes the central stars, however for B1-a these spaxels have low signal. Results might be different for a source with a more homogeneous brightness over a larger part of the field of view.

\begin{table*}
\begin{center}
\caption{Effect of the fringe correction for B1-a.}
\label{tab:fringe correction B1-a}
\small
\begin{tabular}{c|c|c|c|c}
\hline\hline
\multirow{2}{*}{Channel} & \multirow{2}{*}{Band} & \multicolumn{3}{c}{Amplitude (\%)} \\
\cline{3-5}
     & & No correction & Fringe flat & Residual fringe \\
\hline
\multirow{3}{*}{1} & A & $6.3 \;\; [3.91 - 8.62]$ & $3.4 \;\; [1.73 - 5.97]$ & $2.15 \;\; [0.95 - 4.36]$ \\ 
                    & B & $8.02 \;\; [5.46 - 10.85]$ & $3.88 \;\; [1.82 - 6.96]$ & $2.15 \;\; [0.9 - 4.8]$ \\ 
                    & C & $11.09 \;\; [8.15 - 14.87]$ & $4.25 \;\; [2.2 - 8.01]$ & $1.93 \;\; [0.78 - 4.23]$ \\ 
\hline
\multirow{3}{*}{2} & A & $12.45 \;\; [8.46 - 15.97]$ & $4.02 \;\; [2.15 - 7.25]$ & $1.45 \;\; [0.55 - 3.43]$ \\ 
                    & B & $12.58 \;\; [9.17 - 15.61]$ & $2.91 \;\; [1.46 - 4.79]$ & $1.63 \;\; [0.71 - 3.69]$ \\ 
                    & C & $12.23 \;\; [9.83 - 14.65]$ & $2.15 \;\; [1.34 - 3.34]$ & $0.85 \;\; [0.41 - 1.91]$ \\ 
\hline
\multirow{3}{*}{3} & A & $4.16 \;\; [2.67 - 7.36]$ & $1.38 \;\; [0.56 - 2.8]$ & $1.38 \;\; [0.52 - 2.8]$ \\ 
                    & B & $2.79 \;\; [1.34 - 4.13]$ & $0.99 \;\; [0.37 - 1.97]$ & $0.92 \;\; [0.28 - 2.01]$ \\ 
                    & C & $1.7 \;\; [1.05 - 2.7]$ & $0.68 \;\; [0.21 - 1.63]$ & $0.76 \;\; [0.31 - 1.69]$ \\ 
\hline
\multirow{3}{*}{4} & A & $3.6 \;\; [2.21 - 5.61]$ & $1.58 \;\; [0.53 - 3.19]$ & $1.2 \;\; [0.34 - 2.73]$ \\ 
                    & B & $8.11 \;\; [6.11 - 11.01]$ & $2.02 \;\; [0.94 - 4.72]$ & $1.73 \;\; [0.61 - 3.61]$ \\ 
                    & C & $11.89 \;\; [9.33 - 15.16]$ & $2.71 \;\; [1.19 - 5.3]$ & $1.73 \;\; [0.56 - 3.45]$ \\ 
\hline
\hline
\end{tabular}
\normalsize
\end{center}
\textbf{Notes}. See Table \ref{tab:fringe correction NGC-7027} for explanations.
\end{table*}

\section{Discussion and future improvements}
\label{sec:discussion}

\subsection{Fringe correction}

The initial amplitudes of fringes are similar for extended sources and point sources, but they vary with the location on the PSF (detector column) for point sources. As expected, fringe flats derived in this work are efficient for extended sources and reduce fringes to sub-percent levels in most bands. For finite-size sources, especially point-like sources, the fringe pattern is not static but is a function of illumination, which depends sensitively on pointing at the sub-pixel level \citep{Argyriou2020,Gasman2023a}. For these sources, fringes remain at the 1 to 5\% level after calibration by the fringe flat. It depends on the initial fringe amplitude and phase compared to the fringe flat. At long wavelengths (channels 3 and 4), fringes seem to be a combination of an extended source (the background) and of the science object. For semi-extended sources, an analysis of objects with various geometries would be necessary to better characterise fringes. For these sources, improvements of the fringe correction are necessary as the level of noise in the continuum is still larger than expected: the $S/N$ on strong continuum should reach at least 300 if it were limited by photon noise, but is found to be $\sim$80 at most \citep{Beuther2023, Grant2024, Tychoniec2024, vanGelder2024a}, although other limiting factors could also be at play. Point sources that are not located at the same position as point source calibrators also need an improved fringe correction method, for which a library of fringe patterns is being built \citep{Gasman2025}. The residual fringe correction reduces fringes by a factor between one and two for extended sources, and up to a factor three or four for point sources in channels 1 and 2 where it handles the fringe amplitude variations with columns. However, the residual fringe correction can in certain cases reduce the amplitude of periodic astrophysical signatures if these (or the composite spectrum made of multiple individual signatures) happen to have a period similar to that of the detector-induced fringes.

The performance of the fringe flats presented here was evaluated on individual spaxels and, for point and semi-extended sources, on a single dither. For extended sources, the fringe pattern on the detector plane is expected to be the same for all dithers (as evidenced by the fact that coadding images with different pointings produces a well defined fringe pattern). 
For point or semi-extended sources, each dither position has its own fringe pattern, and combining dithers results in a new pattern. How the amplitude estimates given in Tables \ref{tab:fringe correction PDS-70} and \ref{tab:fringe correction B1-a} change after dithers have been combined has yet to be tested (the behaviour is not that of a random noise). As of now, only point source observations with dither positions that are the same as calibrators can be properly corrected for fringes \citep{Pontoppidan2024, Gasman2023a, Gasman2025}.
When dithers are combined and the flux is extracted in an aperture that contains many spaxels (and potentially the background is subtracted), fringes average out to some extent, thus the amplitude of fringes in the final spectrum is reduced compared to the numbers given here. Other noise sources and spectral features from the source become more prominent, thus isolating fringes from these other variations is more challenging in final spectra.
As illustrated in Sect.~\ref{sec:intro}, inaccurate fringe correction can lead to significant line flux uncertainties. Spatially dithered observations with the IFU place a given spectral line at different locations with respect to the fringe pattern. Averaging the extracted spectra lead to an apparently smoother continuum spectrum, but the processes that amplify or reduce the line flux are still at work, and may increase the uncertainty of the measured line flux.
Quantifying this systematic effect for point sources across the MRS wavelength range is non-trivial and requires a dedicated dataset and data analysis that falls outside the scope of this paper. A first attempt to quantify this at a single wavelength (fringe properties change drastically within and across MRS spectral bands) would be to observe a point source that contains a bright emission line in an 8-point dither pattern with target acquisition. The emission line strength should then be compared between the two nominal sets of 4-point dithers. Any deviation in line strength is then linked to detector fringe residuals, as well as uncertainties in the flatfield.

\subsection{Signal-to-noise ratio}
\label{sec:signal-to-noise ratio}

The $S/N$ of the coadded `\texttt{cal}'  images for NGC 7027 and Cat's Eye are shown in Fig.~\ref{fig:SNR}. Here, we estimate the $S/N$ that is required to build fringe flats.
In the fringe flat building process, the columns of each slice were averaged together to fit a single transmittance function for that slice, and the number of columns per slice increases from channel 1 to 4. Also, the fringe period increases in channels 3 and 4 compared to channels 1 and 2 both in wavelength and pixel space (see the frequencies that decrease in Fig. \ref{fig:period-allbands}). Overall, the transmittance function is better sampled for long wavelength bands, which helps the fitting.
To take these effects into account, we computed the number of pixels $N_{\rm{px}}$ that cover half a fringe period (of the main fringe component) weighting each pixel by its amplitude relative to the fringe peak (before any fringe correction) and we defined an effective signal-to-noise ratio $S/N_{\,\rm eff}$ as $S/N\times\sqrt{N_{\rm{px}}}$. This makes $S/N_{\,\rm eff}$ more favourable when the fringes are better sampled. $N_{\rm{px}}$ increases from 50 to 430 from 1A to 4C.

Using Tables \ref{tab:fringe correction NGC-7027} and \ref{tab:fringe correction Cats-Eye}, we found that the fringe peak amplitude is reduced to values below 1\% on individual spaxels for extended sources after fringe flat calibration when $S/N_{\,\rm eff}$ is above 1100. Then, we computed the $S/N$ required to reach that 1\% peak amplitude (that is $1100\,/ \sqrt{N_{\rm{px}}}$) and compared that to the actual $S/N$ given by the \texttt{jwst} pipeline. Fig. \ref{fig:SNR} shows that the actual $S/N$ of NGC~7027 and Cat's~Eye is larger than the required $S/N$ for bands 1C to 4C for NGC 7027 and for bands 3B to 4C for Cat's Eye. This $S/N$ was averaged over the full band; for NGC 7027 the shortest wavelengths of 1B (a third of the band) and the longest wavelengths of 4C (a quarter of the band) are below that limit. Other limitations such as the source spectral lines, pixel response, dichroic fringes, and fringe model accuracy also affect the quality of the fringe flats; nevertheless, these $S/N$ estimates can be useful to design future calibration observations dedicated to building new static fringe flats.
For science observations, the efficiency of the calibration by the static fringe flat should not depend on the $S/N$ of the science source to which it is applied, although subtle effects and their influence on fringes are under investigation \citep[such as the brighter-fatter effect,][]{Argyriou2023b}.

\subsection{Error budget}

Uncertainties in the fringe flats are not formally included in the \texttt{Spec2} pipeline error budget. For bright extended sources observed with MRS, the dominant uncertainty included in the pipeline is that of the flatfield: for NGC 7027, it is the case for bands 1C to 4C, whereas at short wavelengths the read noise and Poisson noise become dominant (1A) or comparable (1B). The flatfield uncertainty on the detector images is between 1 and 2\% for bands 1A to 3C and reaches 4\% in channel 4. 
As shown in Table \ref{tab:fringe correction NGC-7027}, without considering dichroic fringes, the measured fringe peak amplitude after fringe correction is lower (except for bands 1A and 1B where the fringes are not well measured). Thus, the fringe flat uncertainties should not dominate the error budget on individual spaxels and individual dithers for extended sources. They should still be lower when four dithers are averaged, but could become dominant for spectral lines summed over multiple spaxels (because the pixel flat better averages out). In 4B and 4C, after fringe flat and residual fringe correction, the dichroic fringes become dominant and reach amplitudes of several percents in 4C. 
For point sources, the uncertainties on individual dithers and spaxels of the fringe components studied here are comparable to that of the flatfield for channels 1 to 3 and are lower for channel 4; they can become dominant when dithers and spaxels are combined (unless the point source calibrator method is used). Dichroic fringes become dominant in channel 4, in bands 3B and 3C depending on the spaxel, and are generally stronger than for extended sources.
In practice, it is difficult to meaningfully separate the uncertainty of the flatfield from that of the fringe flat (both are full-image corrections divided out of the data), and fringe flat uncertainties are largely incorporated into the flatfield uncertainties. The flatfields may also correct for artefacts and limitations in the fringe flats (for example, the flatfields of the longest wavelength bands contain high frequency variations that could be due to the dichroic fringes).
Improvements of the MRS flatfields are underway (program CAL/MIRI 6619, PI: B. Trahin). After that, an accurate estimate of the fringe flat uncertainties may be necessary. The method developed in Sect. \ref{sec:efficiency} may be used for that purpose but would require improved removal of the continuum and spectral lines (and sampling effect, if possible).

\subsection{Improvements}

Fringe flats would need to be improved at least in bands 1A and 1B using a source that is bright enough at these wavelengths (ground test fringe flats are still in use on CRDS for these two bands at the time of writing). In some other bands, a significant fraction of the wavelength range was excluded because of spectral features of NGC 7027 (up to a third of the band in 1C, 2C, 3A, 4C) which affects the fringe flat accuracy in these regions. Using a bright extended source with less spectral lines would be beneficial.

After applying the fringe flats to NGC 7027, fringes of reduced amplitude remain in most bands, which indicates that the fringe flat does not provide a perfect correction even in this ideal case.
The fringes that remain could be modelled in a second pass using the same procedure as described in Sect. \ref{sec:model}: the initial data would be the NGC 7027 detector images corrected by the fringe flat, and the output would be a second fringe flat that would be merged (multiplied) with the first to produce a finer fringe flat. This idea has not been tested and may work in bands where the residual fringes still follow a regular pattern along the whole band, such as in channel 2 and bands 4B-4C (although these fringes are also reduced by the residual fringe correction).

The dichroic fringes may also be modelled. This third component is not included in the fringe flats because it is very complex (it is a product of the throughput of multiple dichroics in the optical path and is barely resolved in channel 4), it would increase the number of free parameters, and the fitting may become unstable. It could potentially be fitted in a second pass. The fringe model would be different because these fringes originate in different optical elements (the dichroics) and are propagated onto the detector (they do not originate in the detector). They become important at long wavelengths (channels 3 and 4) and their amplitude increases from short to long wavelengths. For point sources, they also depend on the spaxel: some are strongly affected, others are much less affected, even for a similar flux received from the source. The behaviour of the dichroic fringes is different from that of the two fringe components that are currently modelled. 

Other extended sources could be used to build new fringe flats. How much improvement can be expected depends on the spatial homogeneity of the source, its brightness, the total integration time, and the number of spectral features.
Mosaic observations of the Orion Bar were taken on 30 January 2023 (PID~1288, PI: O. Berne). We reduced these data in the same way as NGC 7027, coadding all the images. The surface brightness is lower than NGC 7027 in all channels. It is too faint in channel 1 to improve the fringe flats. In other channels, it provides enough $S/N$, and combining the two data sets would be a way to eliminate some spectral features while preserving data at these wavelengths. The planetary nebula IC 418 was observed on 9 February 2025 to build new flatfields (PID~6619, PI: B. Trahin). This source is more homogeneous than NGC 7027, 1.4 times longer integration times were used, but it is four times fainter in the K band, and the flux should still be too low to derive fringe flats for bands 1A and 1B. A target brighter than NGC 7027 at these wavelengths is necessary for these bands.

Another approach to derive fringe flats would be to use a purely empirical calibration as done for point sources in \citet{Gasman2023a} and \citet{Pontoppidan2024}.
The model we used does not take into account all subtleties (such as the distribution of incidence angles, and only one finesse is allowed to vary), so the model itself is an approximation of the real fringes. The empirical approach consists of extracting the spectrum of a calibrator and using it directly (after continuum removal, spectral line masking, and normalisation) to calibrate the spectrum of other objects. It is done at the column or spaxel level. No modelling is involved. It also removes the dichroic fringes. For point sources, the calibrator can be a star or an asteroid and the target must be placed at the same location on the detector, which can be achieved using target acquisition and the same dither pattern. Extended sources do not have one given location by definition, so such empirical fringe flats could be used directly for other extended sources. 

Conversely, the fringe model used to build fringe flats might be used for semi-extended or point sources, applying the fitting to the source itself. The detector is the same and the initial fringe amplitudes are comparable. The model parameters would depend on the location of the point source with an amplitude and phase that depend on each column. In practice, the fitting procedure is demanding in terms of computing time (and currently works only for extended sources), so adding more parameters may not be feasible, on the other hand the signal is spread over less slices and columns (only a few for point sources). To simplify the procedure, one could search for deviations from a base model. This approach could be useful for semi-extended sources and for point sources that are not placed at the same location as the calibrator. For point sources, other fringe removal methods could also be used \citep[see][for details]{Deming2024}.

Some technical aspects could be improved in the current fringe modelling. It uses spline interpolation to derive parameters along slices with knots equally spaced in wavenumber. The first and last knots are relatively far from the detector edges, and the splines can deviate strongly before the first knot and after the last knot. This can be seen for example in Fig. \ref{fig:finesse-allbands} where the finesse diverges to high or low values near the edges of the band for some slices, which directly impacts the modelled fringe amplitude at these locations. These two knots should be placed closer to the edges. Also, a broad spectral feature from polycyclic aromatic hydrocarbons in NGC 7027 was masked out in band 2C and the fringe flat calibration is not accurate at these wavelengths. The continuum removal seems to work on this feature, so it could be kept. In the last quarter of band 4C ($\lambda>27.5\,\mu$m), the drop in the detector quantum yield results in low recorded fluxes for NGC 7027; most of this region was excluded, nevertheless residual fringes remain in the second half of the spectrum after applying the fringe flat whereas the first half is well corrected. This region could be entirely excluded as it seems to affect the fringe modelling accuracy. It should be noted that discarding spectral features in every band is a manual, tedious process, and choices that are made can only be checked after running the full fringe flat procedure and checking the effect on test data.

Fringe flats are independent of wavelength calibration and should not need to change with pipeline version. 
The fringe flat building and calibration are done before the spectrophotometric calibration (both calibrations are multiplicative pixel by pixel) so fringe flats are not affected by flux calibration uncertainties or time-dependencies.
The current fringe flats were built from the `\texttt{rate}'  files, thus without flatfield calibration. Using flatfield calibrated images may help refine the fringe flats by reducing the dispersion of the data points before fitting the fringe model. In practice, the fringe flats were built first, and care was taken to remove fringes when building the flatfields.

\section{Conclusion}
\label{sec:conclusion}

Correcting for interference fringes in the JWST MIRI Medium Resolution Spectrometer is crucial to detect weak spectral features on a strong continuum and measure accurate line fluxes and line-flux ratios. Extended source fringe flats for MRS have been presented in this paper. Fringe flats were built from observations of the planetary nebula NGC 7027 for bands 1C to 4C while those obtained from ground tests are still in use for bands 1A and 1B. These fringe flats are available on CRDS and are used by the \texttt{fringe} step of the \verb|Spec2Pipeline| of the JWST Science Calibration Pipeline.
The fringe model included two fringe components that originate in the detector. One dominates at short wavelengths (channel 1), the other at long wavelengths (channel 4), and they beat against each other at middle wavelengths (channels 2 and 3). 
A third fringe component of higher frequency that originates from the dichroic filters appears in channels 3 and 4. It was not included in the fringe model and thus is not corrected for by the fringe flats.
The performance of the fringe flat calibration has been evaluated on sources of different types (extended, semi-extended, and point sources) using individual spaxels and, for point and semi-extended sources, a single dither position (the dither position should not matter for extended sources).
As expected, the fringe flats perform well on extended sources and reduce fringes to the sub-percent level, but are less efficient for point-sources and the semi-extended source studied here (that consists mostly in two point sources). The 2D residual fringe correction further decreases fringes: it is most useful for point sources in bands where the fringe amplitude depends on the column, whereas its effect is small and noticeable only in some bands for extended sources (to the level of precision that can be measured with our method). The fringe amplitudes reported in Tables \ref{tab:fringe correction NGC-7027} to \ref{tab:fringe correction B1-a} are expected to be similar for other dither positions, and to decrease when combining several dithers, summing many spaxels to extract science spectra, and applying the 1D residual fringe correction.

Future improvements of the fringe flats are possible. Flight fringe flats should be built for bands 1A and 1B from an extended source that is bright enough at these short wavelengths. An extended source that is bright at the longest wavelengths of 4C would also be useful. A bright source with less spectral features could be used to improve the fringe flats in other bands, and combining data from several extended sources could help. A second pass of the fringe modelling may decrease residual fringes, and the model could be updated to include the dichroic fringes. A purely empirical approach could also be investigated for extended sources, as done for point sources. The detector physical properties responsible for fringes are not expected to change over time, and thus the fringe flats should not need regular updates.

\begin{acknowledgements}

IA thanks the European Space Agency (ESA) and the Belgian Federal Science Policy Office (BELSPO) for their support in the framework of the PRODEX Programme.
EvD acknowledges support from A-ERC grant 101019751 MOLDISK.
The following National and International Funding Agencies funded and supported the MIRI development: NASA; ESA; Belgian Science Policy Office (BELSPO); Centre National d'Etudes Spatiales (CNES); Danish National Space Centre; Deutsches Zentrum für Luft- und Raumfahrt (DLR); Enterprise Ireland; Ministerio De Economía y Competividad; Netherlands Research School for Astronomy (NOVA); Netherlands Organisation for Scientific Research (NWO); Science and Technology Facilities Council; Swiss Space Office; Swedish National Space Agency; and UK Space Agency.
MIRI draws on the scientific and technical expertise of the following organizations: Ames Research Center, USA; Airbus Defence and Space, UK; CEA-Irfu, Saclay, France; Centre Spatial de Liège, Belgium; Consejo Superior de Investigaciones Científicas, Spain; Carl Zeiss Optronics, Germany; Chalmers University of Technology, Sweden; DTU Space Technical University of Denmark; Dublin Institute for Advanced Studies,
Ireland; European Space Agency, Netherlands; ETCA, Belgium; ETH Zurich, Switzerland; Goddard Space Flight Center, USA; Institut d'Astrophysique Spatiale, France; Instituto Nacional de Técnica Aeroespacial, Spain; Institute for Astronomy, Edinburgh, UK; Jet Propulsion Laboratory, USA; Laboratoire d'Astrophysique de Marseille (LAM), France; Leiden University, Netherlands; Lockheed Advanced Technology Center, USA; NOVA Opt-IR group at Dwingeloo, Netherlands; Northrop Grumman, USA; Max Planck Institut für Astronomie (MPIA), Heidelberg, Germany; Laboratoire d'Etudes Spatiales et d'Instrumentation en Astrophysique (LESIA), France; Paul Scherrer Institut, Switzerland; Raytheon Vision Systems, USA; RUAG Aerospace, Switzerland; Rutherford Appleton Laboratory (RAL Space), UK; Space Telescope Science Institute, USA; Toegepast-Natuurwetenschappelijk Onderzoek (TNO-TPD), Netherlands; UK Astronomy Technology Centre, UK; University College London, UK; University of Amsterdam, Netherlands; University of Arizona, USA; University of Bern, Switzerland; University of Cardiff, UK; University of Cologne, Germany; University of Ghent, Belgium; University of Groningen, Netherlands; University of Leicester, UK; University of Leuven, Belgium; Stockholm University, Sweden; and Utah State University, USA.
This work is based on observations made with the NASA/ESA/CSA James Webb Space Telescope. The data were obtained from the Mikulski Archive for Space Telescopes at the Space Telescope Science Institute, which is operated by the Association of Universities for Research in Astronomy, Inc., under NASA contract NAS 5-03127 for JWST. These observations are associated with programs \# 1031, 1523, 1282, 1549, 1290, and 1288.
The authors acknowledge the team led by PI D. R. Law, the team led by PI K. M. Pontoppidan, and the team led by coPIs O. N. Berne, E. Peeters, and E. Habart for developing their observing programs with a zero-exclusive-access period.

\end{acknowledgements}

\bibliographystyle{aa}

\end{document}